\documentclass[]{aa}
\usepackage{psfig}
\usepackage{times}

\begin{document}
\newcommand{\pdrv}[2]{\frac{\partial #1}{\partial #2}}
\newcommand{\drv}[2]{{{{\rm d} #1}\over {{\rm d} #2}}}

   \thesaurus{06 	    
              (08.05.3;     
               08.13.2;     
	       08.02.3;     
               08.14.1;     
               08.23.1;     
               03.13.4)}    
 
   \title{Formation of millisecond pulsars}
   \subtitle{I. Evolution of low-mass {X}-ray binaries with 
             $P_{\rm orb}>$ 2 days}

   \author{Thomas M. Tauris \inst{} \& Gerrit J. Savonije}

   \offprints{tauris@astro.uva.nl}

   \institute{Center for High-Energy Astrophysics,
              University of Amsterdam,
              Kruislaan 403, NL-1098 SJ Amsterdam, The Netherlands}

   \date{Received 26 April 1999 / Accepted 12 August 1999}

   \maketitle \markboth{T.M. Tauris \& G.J. Savonije: Formation of
                        millisecond pulsars. I}{}

\begin{abstract} 
We have performed detailed numerical calculations of the non-conservative
evolution of close binary systems with low-mass ($1.0-2.0\,M_{\odot}$) donor
stars and a $1.3\,M_{\odot}$ accreting neutron star.
Rather than using analytical expressions for simple polytropes, we calculated
the thermal response of the donor star to mass loss, in order to
determine the stability and follow the evolution of the mass transfer.
Tidal spin-orbit interactions and Reimers wind mass-loss were also
taken into account.\\
We have re-calculated the correlation between orbital period and white dwarf
mass in wide binary radio pulsar systems. Furthermore, we find an 
anti-correlation
between orbital period and neutron star mass under the assumption of the
"isotropic re-emission" model and compare this result with observations.
We conclude that the accretion
efficiency of neutron stars is rather low and that they eject a
substantial fraction of the transferred material even when accreting
at a sub-Eddington level. 

The mass-transfer rate is a strongly increasing function of
initial orbital period and donor star mass.
For relatively close systems with light donors ($P_{\rm orb}<10$ days
and $M_2 <1.3\,M_{\odot}$) the mass-transfer rate is sub-Eddington,
whereas it can be highly super-Eddington by a factor of $\sim\!10^4$ for
wide systems with relatively heavy donor stars ($1.6\sim 2.0\,M_{\odot}$)
as a result of their deep convective envelopes. 
We briefly discuss the evolution of {X}-ray binaries with donor stars
in excess of $2\,M_{\odot}$.

Based on our calculations we present evidence that
PSR J1603--7202 evolved through a phase with unstable mass transfer from
a relatively heavy donor star and therefore is likely to host a
{CO} white dwarf companion.

\end{abstract}

\section{Introduction}
Millisecond pulsars are characterized by short rotational periods
($P_{\rm spin}<30$ ms) and relatively weak surface magnetic fields ($B<10^{10}$ 
G)
and are often found in binaries with a white dwarf companion.
They are old neutron stars which have been recycled in a close binary via
accretion of mass and angular momentum from a donor star.
The general scenario of this process is fairly well understood qualitatively
(cf. review by Bhattacharya \& van~den~Heuvel 1991), but there
remain many details which are still uncertain and difficult to
analyze quantitatively. It is our aim to highlight these problems in a
series of papers and try to answer them using detailed numerical
calculations with refined stellar evolution and binary interactions.

There are now more than 30 binary millisecond pulsars known in the Galactic
disk. They can be roughly divided into three observational classes (Tauris 
1996).
Class$\,${A} contains the wide-orbit ($P_{\rm orb}> 20$ days)
binary millisecond pulsars (BMSPs)
with low-mass helium white dwarf companions ($M_{\rm WD} < 0.45\,M_{\odot}$),
whereas the close-orbit BMSPs ($P_{\rm orb}\la 15$ days) consist of systems
with either low-mass helium white dwarf companions (class$\,${B}) or systems
with relatively heavy {CO} white dwarf companions (class$\,${C}).
The latter class evolved through a phase with significant loss of 
angular momentum (e.g. common envelope evolution)
and descends from systems with a heavy donor star: $2 < M_2/M_{\odot} < 6$.
The single millisecond pulsars
are believed to originate from tight class$\,${B} systems where the
companion has been destroyed or evaporated 
-- either from X-ray irradiation when the
neutron star was accreting, or in the form of a pulsar radiation/wind
of relativistic particles (e.g. Podsiadlowski 1991; Tavani 1992).

The evolution of a binary initially consisting of a neutron star and a 
main-sequence companion depends on the mass of the companion (donor) star 
and the initial orbital period of the system. If the donor star is 
heavy compared to the neutron star then the mass transfer is likely to result
in a common envelope (CE) evolution (Paczynski 1976; Webbink 1984; Iben \& Livio
1993) where the neutron star spirals in through the envelope of the donor
in a very short timescale of less than $10^4$ yr. 
The observational paucity of Roche-lobe filling companions more
massive than $\sim\!2\,M_{\odot}$ has been attributed
to their inability to transfer mass in a stable mode such
that the system becomes a persistent long-lived X-ray source
(van~den~Heuvel 1975; Kalogera \& Webbink 1996).
For lighter donor stars ($< 2\,M_{\odot}$) the system evolves into a
low-mass X-ray binary (LMXB) which evolves on a much longer timescale
of $10^7-10^9$ yr. It has been shown by Pylyser \& Savonije
(1988,1989) that an orbital bifurcation period ($P_{\rm bif}$)
separates the formation of converging systems (which evolve with
decreasing orbital periods until the mass-losing component becomes
degenerate and an ultra-compact binary is formed) from the diverging
systems (which finally evolve with increasing orbital periods until the
mass losing star has lost its envelope and a wide detached binary is formed).
It is the LMXBs with $P_{\rm orb}>P_{\rm bif}$ ($\simeq 2$ days) which are
the subject of this paper -- the progenitors of the wide-orbit 
class$\,${A} BMSPs.

In these systems the mass transfer is driven by the interior 
thermonuclear evolution
of the companion star since it evolves into a (sub)giant before loss of
orbital angular momentum dominates.
In this case we get an LMXB with a giant donor.
These systems have been studied by 
Webbink, Rappaport \& Savonije (1983), Taam (1983), Savonije (1987),
Joss, Rappaport \& Lewis (1987)
and recently Rappaport~et~al. (1995) and Ergma, Sarna \& Antipova (1998).
For a donor star on the red giant branch (RGB) the growth in core-mass 
is directly related to the luminosity,
as this luminosity is entirely generated by hydrogen shell burning.
As such a star, with a small compact core surrounded by en extended
convective envelope, is forced to move up the Hayashi track its luminosity
increases strongly with only a fairly modest decrease in temperature.
Hence one also finds a relationship between the giant's radius and the mass
of its degenerate helium core -- almost entirely independent of the mass 
present in the hydrogen-rich envelope (Refsdal \& Weigert 1971;
Webbink, Rappaport \& Savonije 1983). 
In the scenario under consideration, the extended envelope of the giant
is expected to fill its Roche-lobe until termination of the mass transfer.
Since the Roche-lobe radius $R_{\rm L}$ only depends on the masses
and separation between the two stars it is clear that the core-mass, from
the moment the star begins Roche-lobe overflow, is uniquely correlated with 
the orbital period of the system.
Thus also the final orbital period, $P_{\rm orb}^{\rm f}$ ($2\sim 10^{3}$ days)
is expected to be a function of the mass
of the resulting white dwarf companion (Savonije 1987).
It has also been argued that the core-mass determines the rate of mass transfer
(Webbink, Rappaport \& Savonije 1983).
For a general overview of the evolution of LMXBs -- see e.g. Verbunt (1990).\\
In this study we also discuss the final post-accretion mass of the 
neutron star and confront it with observations and  
the consequences of the new theory for kaon 
condensation in the core of neutron stars which result in a very soft
equation-of-state and a corresponding maximum neutron star mass
of only $\sim\!1.5 M_{\odot}$ (Brown \& Bethe 1994).

In Section~2 we briefly introduce the code, and in Sections~3 and 4
we outline the orbital evolution and the stability criteria for mass transfer.
We present the results of our LMXB calculations in Section~5 and
in Section~6 we discuss our results and compare with observations.
Our conclusions are given in Section~7 and a summary table of
our numerical calculations is presented in the Appendix.

\section{A brief introduction to the numerical computer code}
We have used an updated version of the numerical stellar evolution code 
of Eggleton. This code uses a self-adaptive, non-Lagrangian mesh-spacing 
which is a function of local pressure, temperature, Lagrangian mass and radius.
It treats both convective and semi-convective mixing as a diffusion process
and finds a simultaneous and implicit solution of both the stellar
structure equations and the diffusion equations for the chemical composition.
New improvements are the inclusion of pressure ionization and
Coulomb interactions in the equation-of-state, and the incorporation of recent
opacity tables, nuclear reaction rates and neutrino loss rates.
The most important recent updates of this code are described in
Pols~et~al. (1995;$\,$1998) and some are summarized in
Han, Podsiadlowski \& Eggleton (1994).\\
We performed such detailed numerical stellar evolution calculations in our work
since they should result in more realistic results compared to
models based on complete, composite or condensed polytropes.

We have included a number of binary interactions in this code
in order to carefully follow the details of the mass-transfer process in LMXBs.
These interactions include losses of orbital angular momentum due to
mass loss, magnetic braking, gravitational wave radiation and the effects of
tidal interactions and irradiation of the donor star by hard photons 
from the accreting neutron star.

\section{The equations governing orbital evolution}
The orbital angular momentum for a circular\footnote{We assume 
circular orbits throughout this paper -- tidal effects
acting on the near RLO giant star will circularize the orbit anyway
on a short timescale of $\sim\!10^4$ yr, cf. Verbunt \& Phinney (1995).}
binary is:
\begin{equation}
  J_{\rm orb} = \frac{M_{\rm NS}\,M_2}{M}\,\Omega\,a^2
\end{equation}
where $a$ is the separation between the stellar components,
$M_{\rm NS}$ and $M_2$ are the masses of the (accreting) neutron star
and the companion (donor) star, respectively,
$M=M_{\rm NS}+M_2$ and the orbital angular velocity,
$\Omega = \sqrt{GM/a^3}$. Here $G$ is the constant of gravity.
A simple logarithmic differentiation of this equation yields
the rate of change in orbital separation:
\begin{equation}
  \frac{\dot{a}}{a} = 2\frac{\dot{J}_{\rm orb}}{J_{\rm orb}}
                     -2\frac{\dot{M}_{\rm NS}}{M_{\rm NS}}
                     -2\frac{\dot{M}_2}{M_2}
                     +\frac{\dot{M}_{\rm NS}+\dot{M}_2}{M}
\end{equation}
where the total change in orbital angular momentum is:
\begin{equation}
 \frac{\dot{J}_{\rm orb}}{J_{\rm orb}} =
  \frac{\dot{J}_{\rm gwr}}{J_{\rm orb}} + \frac{\dot{J}_{\rm mb}}{J_{\rm orb}}
  +\frac{\dot{J}_{\rm ls}}{J_{\rm orb}} + \frac{\dot{J}_{\rm ml}}{J_{\rm orb}}
\end{equation}
The first term on the right side of this equation gives
the change in orbital angular momentum due to gravitational wave
radiation (Landau \& Lifshitz 1958):
\begin{equation}
  \frac{\dot{J}_{\rm gwr}}{J_{\rm orb}} = - \frac{32\,G^{3}}{5\,c^{5}}
  \frac{M_{\rm NS}\,M_2\,M}{a^4} \qquad \mbox{${\rm s}^{-1}$}
\end{equation}
where $c$ is the speed of light in vacuum.
The second term arises due to magnetic braking.  This is is a combined effect of
a magnetically coupled stellar wind and tidal spin-orbit coupling which tend to
keep the donor star spinning synchronously with the orbital motion.
Observations of low-mass dwarf stars with rotational periods in the range of
$1\sim 30$ days (Skumanich 1972) show that even a weak (solar-like) wind will
slow down their rotation in the course of time due to interaction of the stellar
wind with the magnetic field induced by the differential rotation in the
convective envelope.  For a star in a close binary system, the rotational
braking is compensated by tidal coupling so that orbital angular momentum is
converted into spin angular momentum and the binary orbit shrinks.  Based on
this observed braking law correlation between rotational period and age, Verbunt
\& Zwaan (1981) estimated the braking torque and we find:  
\begin{equation}
  \frac{\dot{J}_{\rm mb}}{J_{\rm orb}}\approx -0.5\times 10^{\rm -28}\,f_{\rm mb}^{-2}
  \;\frac{IR_2^2}{a^5}\;\frac{GM^2}{M_{\rm NS}\,M_2} \qquad\mbox{${\rm s}^{-1}$}
\end{equation} 
where $R_2$ is the radius of the donor star, $I$ is its moment of inertia and
$f_{\rm mb}$ is a constant of order unity (see also discussion by
Rappaport, Verbunt \& Joss 1983). In order to sustain a
significant surface magnetic field we required a minimum depth of $Z_{\rm
conv}>0.065\,R_{\odot}$ for the convective envelope (cf.  Pylyser \& Savonije
1988 and references therein).  Since the magnetic field is believed to be
anchored in the underlaying radiative layers of the star (Parker 1955), we also
required a maximum depth of the convection zone:  $Z_{\rm conv}/R_2 < 0.80$ in
order for the process of magnetic braking to operate.  These limits imply that
magnetic braking operates in low-mass ($M_2 \la 1.5\,M_{\odot}$) stars which are
not too evolved.

The third term on the right side of eq.~(3) describes possible exchange of
angular momentum between the orbit and the donor star due to its expansion or
contraction. For both this term and the magnetic braking term we estimate
whether or not the tidal torque is sufficiently strong to keep the donor star
synchronized with the orbit. 
The tidal torque is determined by considering the effect of turbulent
viscosity in the convective envelope
of the donor on the equilibrium tide. 
When the donor star approaches
its Roche-lobe tidal effects become strong and lead to synchronous
rotation.
The corresponding tidal energy dissipation rate
was calculated and taken into account in the local energy balance of the star.
The tidal dissipation term was distributed through the convective envelope
according to the local mixing-length approximation for turbulent
convection -- see Appendix for further details.

Since we present calculations here for systems with $P_{\rm orb}>2$ days,
the most significant contribution to the overall change in orbital
angular momentum is caused by loss of mass from the system.
This effect is given by:
\begin{equation}
  \frac{\dot{J}_{\rm ml}}{J_{\rm orb}} \approx
  \frac{\alpha+\beta\,q^2+\delta\,\gamma\,(1+q)^2}{1+q}\;
  \frac{\dot{M}_2}{M_2}
\end{equation}
Here $q\equiv M_2/M_{\rm NS}$ is the mass ratio of the donor over the accretor 
and $\alpha$, $\beta$ and $\delta$ are the fractions of mass lost from the donor
in the form of a fast wind, the mass ejected from the vicinity of the
neutron star and from a circumstellar coplanar toroid (with
radius, $a_{\rm r}=\gamma^2\,a$), respectively  -- see van~den~Heuvel (1994a)
and Soberman, Phinney \& van~den~Heuvel (1997).
The accretion efficiency of the neutron star is thus given by:
$\epsilon = 1 -\alpha -\beta -\delta$, or equivalently:
\begin{equation}
  \partial M_{\rm NS} = -(1-\alpha-\beta-\delta)\;\partial M_2
\end{equation}
where $\partial M_2 < 0$. Note, that these factors will also be
functions of time as the binary evolve. 
Low-mass ($1-2\,M_{\odot}$) donor stars
do not lose any significant amount material in the form of a direct wind
-- except for an irradiated donor in a very close binary system,
or an extended giant donor evolving toward the tip of the RGB
which loses a significant amount of material in a wind. 
For the latter type of donors we used Reimers' (1975) formula to
calculate the wind mass-loss rate:
\begin{equation}
  \dot{M}_{2\,\rm wind} = -4\times 10^{-13}\;\eta_{\rm RW}\,
        \frac{L\,R_2}{M_2} 
        \qquad \mbox{$M_{\odot}$ yr$^{-1}$}
\end{equation}
where the mass, radius and luminosity are in solar units and
$\eta_{\rm RW}$ is the mass-loss parameter. We assumed
$\eta_{\rm RW}=0.5$ for our work  -- cf. Renzini (1981) and
Sackmann, Boothroyd \& Kraemer (1993) for discussions.
The mass-loss mechanism involving a circumstellar toroid
drains far too much orbital angular momentum from the LMXB 
and would be dynamical unstable resulting in
a runaway event and formation of a CE.
Also the existence of binary radio pulsars with orbital
periods of several hundred days exclude this scenario as being dominant.\\
Hence, for most of the work in this paper we have
$\alpha \ll \beta$, and we shall
assume $\delta =0$, and for LMXBs with large mass-transfer
rates the mode of mass transfer to consider is therefore the
"isotropic re-emission" model. In this model
all of the matter flows over, in a conservative
way, from the donor star to an accretion disk in the vicinity of the neutron 
star, and then a fraction, $\beta$ of this material is ejected isotropically
from the system with the specific orbital angular momentum of the neutron star.

As mentioned above, since we present calculations here for systems with
initial periods larger than 2 days, loss of angular momentum due
to gravitational wave radiation and magnetic braking (requiring
orbital synchronization) will in general not be very significant.

\subsection{The mass-transfer rate}
For every timestep in the evolution calculation of the donor star the 
mass-transfer rate is calculated from the boundary condition on the 
stellar mass:
\begin{equation}
 \dot{M}_2=-1 \times 10^{3}\;
             PS \left[\ln{\frac{R_2}{R_{\rm L}}}\right]^3
        \qquad \mbox{$M_{\odot}$ yr$^{-1}$}
\end{equation}
where $PS[x]=0.5\,[x+abs(x)]$ and $R_{\rm L}$ is the donor's
Roche-radius given by (Eggleton 1983):  
\begin{equation} 
  R_{\rm L} = \frac{0.49\,q^{2/3}\:a}{0.6\,q^{2/3}+\ln(1+q^{1/3})}
\end{equation}
The orbital separation $a$ follows from the orbital angular momentum 
balance -- see eqs~(1) and (3). All these variables are included in a Henyey 
iteration scheme. The above expression for the mass-transfer rate is rather
arbitrary, as is the precise amount of Roche-lobe overfill for a certain 
transfer rate; but the results are independent of the precise form as they are 
determined by the response of the stellar radius to mass loss. 

\section{Stability criteria for mass transfer}
The stability and nature of the mass transfer is very important in binary 
stellar
evolution.  It depends on the response of the mass-losing donor star and of the
Roche-lobe -- see Soberman, Phinney \& van~den~Heuvel (1997) for a nice review.
If the mass transfer proceeds on a short timescale (thermal or dynamical) the
system is unlikely to be observed during this short phase, whereas if the mass
transfer proceeds on a nuclear timescale it is still able to sustain a high 
enough
accretion rate onto the neutron star for the system to be observable as an LMXB
for an appreciable interval of time.

When the donor evolves to fill its Roche-lobe (or alternatively, the binary
shrinks sufficiently as a result of orbital angular momentum losses) the
unbalanced pressure at the first Lagrangian point will initiate mass transfer
(Roche-lobe overflow, RLO) onto the neutron star.  When the donor star is
perturbed by removal of some mass, it falls out of hydrostatic and thermal
equilibrium.  In the process of re-establishing equilibrium, the star will 
either
grow or shrink -- first on a dynamical (sound crossing), and then on a slower
thermal (heat diffusion, or Kelvin-helmholtz) timescale.  Also the Roche-lobe
changes in response to the mass transfer/loss.  As long as the donor star's
Roche-lobe continues to enclose the star the mass transfer is stable. Otherwise 
it is unstable and proceeds on a dynamical timescale.  Hence the question of
stability is determined by a comparison of the exponents in power-law fits of
radius to mass, $R\propto M^{\zeta}$, for the donor star and the Roche-lobe,
respectively:
\begin{equation}
  \zeta_{\rm donor} \equiv \frac{\partial\ln R_2}{\partial\ln M_2}
                       \;\;\;   \wedge \;\;\;
  \zeta_{\rm L} \equiv \frac{\partial\ln R_{\rm L}}{\partial\ln M_2}
\end{equation}
where $R_2$ and $M_2$ refer to the mass losing donor star.
Given $R_2=R_{\rm L}$ (the condition at the onset of RLO) 
the initial stability criterion becomes:
\begin{equation}
  \zeta_{\rm L} \le \zeta_{\rm donor}
\end{equation}
where $\zeta_{\rm donor}$ is the adiabatic or thermal (or somewhere
in between) response of the donor star to mass loss. Note,
that the stability might change during
the mass-transfer phase so that initially stable systems become
unstable, or vice versa, later in the evolution.
The radius of the donor is a function of time and mass and thus:
\begin{equation}
  \dot{R}_2 = \left.\frac{\partial R_2}{\partial t}\right|_{\rm M_2}
              + R_2\,\zeta_{\rm donor}\,\frac{\dot{M}_2}{M_2}
\end{equation}
\begin{equation}
  \dot{R}_{\rm L}=\left.\frac{\partial R_{\rm L}}{\partial t}\right|_{\rm M_2}
              + R_{\rm L}\,\zeta_{\rm L}\,\frac{\dot{M}_2}{M_2}
\end{equation}
The second terms follow from eq.~(11); the first term of eq.~(13)
is due to expansion of the donor star as a result of nuclear
burning (e.g. shell hydrogen burning on the RGB) and the first term
of eq.~(14) represents changes in $R_{\rm L}$ which are not caused
by mass transfer such as orbital decay due to gravitational
wave radiation and tidal spin-orbit coupling. Tidal coupling
tries to synchronize the orbit whenever the rotation of the donor
is perturbed (e.g. as a result of magnetic braking or an increase of 
the moment of inertia while the donor expands).
The mass-loss rate of the donor can be found as a self-consistent
solution to eqs~(13) and (14) assuming $\dot{R}_2 = \dot{R}_{\rm L}$
for stable mass transfer.

\subsection{The Roche-radius exponent, $\zeta_{\rm L}$}
For binaries with orbital periods larger than a few days it is a
good approximation that $\dot{J}_{\rm gwr}, \dot{J}_{\rm mb} \ll \dot{J}_{\rm ml}$
and $\alpha \ll \beta$ during the RLO mass-transfer phase.
Assuming $\dot{J}_{\rm gwr}\!=\!\dot{J}_{\rm mb}\!=\!0$
and $\alpha\!=\!\delta\!=\!0$
we can therefore use the analytical expression obtained by
Tauris (1996) for an integration of eq.~(2)
to calculate the change in orbital separation
during the LMXB phase (assuming a constant $\beta$):
\begin{equation}
  \frac {a}{a_0} = \left(\frac{q_0\,(1-\beta)+1}
        {q\,(1-\beta)+1}\right)^{\textstyle\frac{3\beta -5}{1-\beta}}
        \left(\frac{q_0+1}{q+1}\right)\,
        \left(\frac{q_0}{q}\right)^{2}\;\Gamma_{\rm ls}
\end{equation}
where the subscript `0' denotes initial values.
Here we have added an extra factor, $\Gamma_{\rm ls}$:
\begin{equation}
  \Gamma_{\rm ls} = \exp\,
                    \left({2\int_{0}\frac{(dJ)_{\rm ls}}{J_{\rm orb}}}\right)
\end{equation}
to account for the tidal spin-orbit coupling since
$\dot{J}_{\rm ls}\neq0$.
One aim of this study is to evaluate the
deviation of $\Gamma_{\rm ls}$ from unity.

If we combine eqs~(7), (10) and (15), assuming $\Gamma_{\rm ls}=1$,
we obtain analytically:
\begin{eqnarray}
  \zeta_{\rm L}&=&\frac{\partial \ln R_{\rm L}}{\partial \ln M_2} =
                 \left( \frac{\partial \ln a}{\partial \ln q} +
                        \frac{\partial \ln (R_{\rm L}/a)}{\partial \ln q}
                 \right) \: \frac{\partial \ln q}{\partial \ln M_2} \nonumber\\
                                                                    \nonumber\\
               &=& [1+(1-\beta)\,q]\,\psi + (5-3\beta)\,q
\end{eqnarray}
where
\begin{equation}
  \psi = \left[ -\frac{4}{3} -\frac{q}{1+q} -
                \frac{2/5 + 1/3\,q^{-1/3}\,(1+q^{1/3})^{-1}}
                                    {0.6 + q^{-2/3}\ln(1+q^{1/3})} \right]
\end{equation}
In the limiting case where $q\rightarrow 0$ (when the accretor is
much heavier than the donor star) we find:
\begin{equation}
  \lim_{q \rightarrow 0}\; \zeta_{\rm L} = - 5/3
\end{equation}
The behavior of $\zeta_{\rm L}\,(q,\beta)$ for LMXBs is shown in
Fig.~1. We note that $\zeta_{\rm L}$ does not depend strongly on $\beta$.
This figure is quite useful to get an idea of the stability of a given
mass transfer when comparing with $\zeta$ for the donor star. We see that
in general the Roche-lobe, $R_{\rm L}$ increases 
($\zeta_{\rm L}<0$) when material
is transferred from a light donor to a heavier NS ($q<1$) and
correspondingly $R_{\rm L}$ decreases ($\zeta_{\rm L}>0$) when material
is transferred from a heavier donor to a lighter NS ($q>1$).
This behavior is easily understood from the bottom panel of 
the same figure where we have plotted $-\partial\ln(a)/\partial\ln(q)$
as a function of $q$. The sign of this quantity is important since
it tells whether the orbit expands or contracts in response to
mass transfer (note $\partial q<0$). 
We notice that the orbit always expands when $q<1$
and it always decreases when $q> 1.28$,
whereas for $1 < q \la 1.28$ it can still expand if $\beta >0$. 
There is a point at $q=3/2$ where $\partial\ln(a)/\partial\ln(q)=2/5$
is independent of $\beta$.
It should be mentioned that if $\beta >0$ then, in some cases,
it is actually possible to decrease
the separation, $a$ between two stellar components while
increasing $P_{\rm orb}$ at the same time!
  \begin{figure}[t]
    \psfig{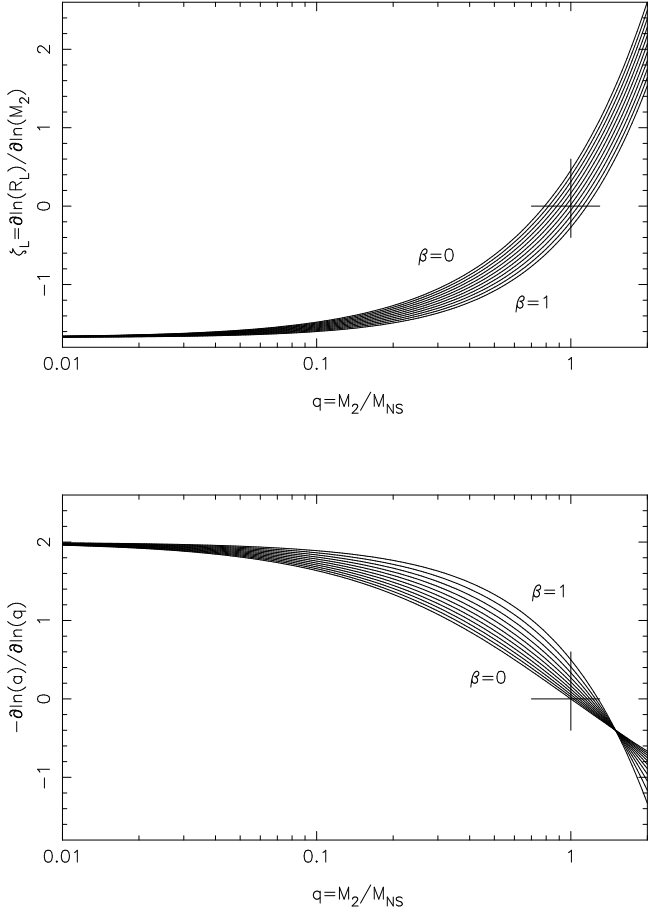}
     \caption{Top panel: the Roche-radius exponent
             ($R_{\rm L}\propto M_2^{\zeta_{\rm L}}$)
             for LMXBs as a function of $q$ and $\beta$.
             The different curves correspond to different
             constant values of $\beta$ in steps of 0.1.
             Tidal effects were not taken into account 
             ($\Gamma_{\rm ls}=1$). A cross is shown to highlight
             the case of $q=1$ or $\zeta_{\rm L}=0$.
             In the bottom panel we have plotted 
             $-\partial\ln(a)/\partial\ln(q)$ as a function of $q$.
             The evolution during the mass-transfer phase follows
             these curves from right to left (though $\beta$ need
             not be constant) since $M_2$ and $q$ are decreasing
             with time. See text for further explanation.
             }
  \end{figure}

\section{Results}
We have evolved a total of a few hundred different LMXB systems.
121 of these are listed in Table~A1 in the Appendix.
We chose donor star masses of $1.0\leq M_2/M_{\odot} \leq 2.0$
and initial orbital periods of $2.0\la P_{\rm orb}^{\rm ZAMS}/\rm{days} \leq 
800$.
We also evolved donors with different chemical compositions and
mixing-length parameters. In all cases we assumed an initial neutron star
mass of $1.3\,M_{\odot}$.\\
In Fig.~2 we show the evolution of four LMXBs.
As a function of donor star mass
($M_2$) or its age since the ZAMS, we have plotted the 
orbital period ($P_{\rm orb}$), the mass-loss rate of the donor
as well as the mass-loss rate from the system ($\dot{M}_2$ and
$\dot{M}$), the radius exponent ($\zeta$) of the donor and its Roche-lobe
and finally the depth of the donor's convection zone ($Z_{\rm conv}/R_2$).
Note, that we have zoomed in on the age interval which corresponds
to the mass-transfer phase.
As an example, we have chosen two different initial donor masses 
($1.0\,M_{\odot}$ and $1.6\,M_{\odot}$) -- 
each with two different initial orbital periods (3.0 and 60.0 days)
of the neutron star (NS) and its ZAMS companion.
The evolutionary tracks of the donor stars are plotted in the
HR-diagram in Fig.~3.
We will now discuss the evolution of each of these systems in more detail.
  \begin{figure*}
    \psfig{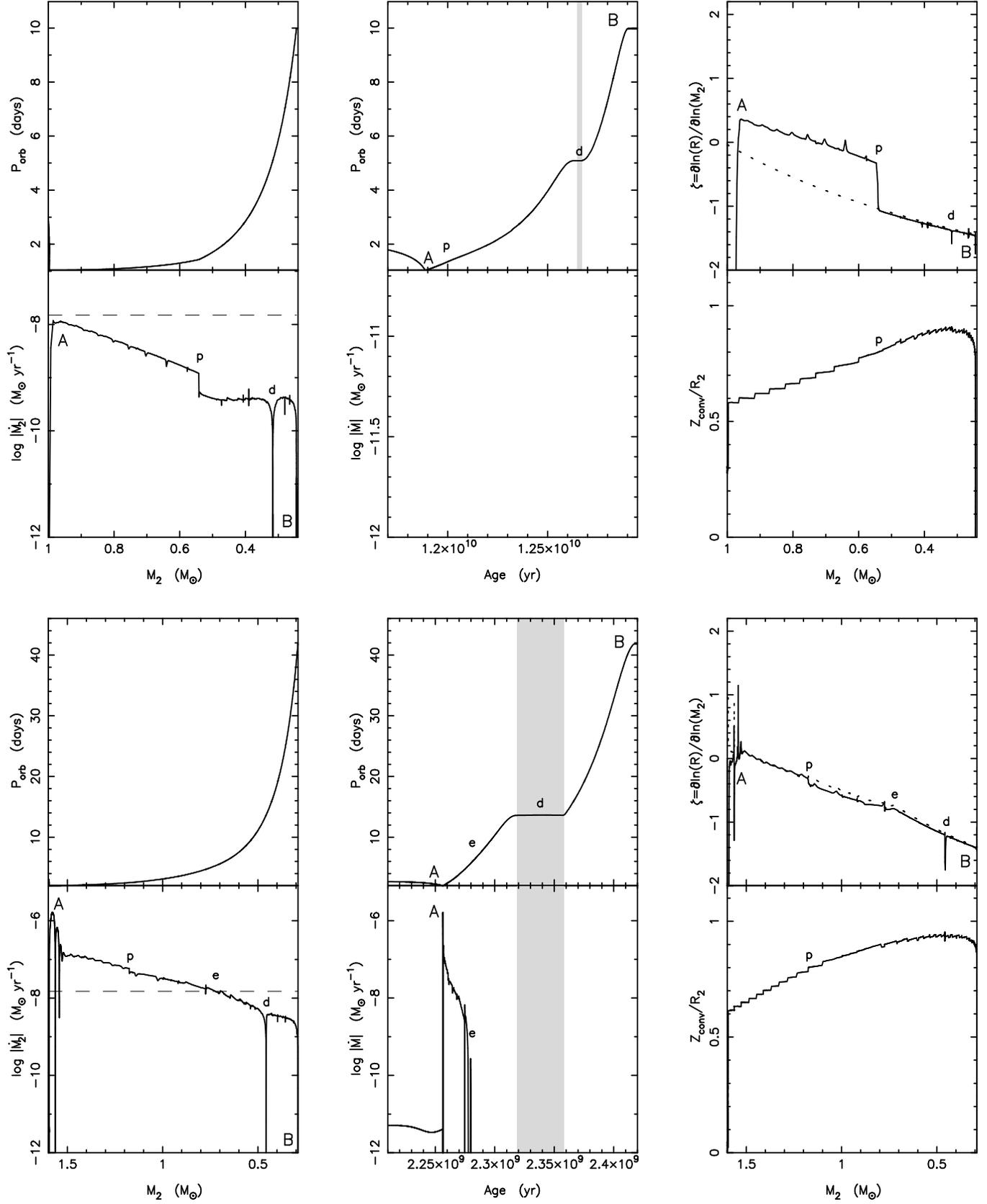}
    \caption[]{\hspace{-0.16cm}{\bf a-b.}\hspace{0.16cm}
               Numerical calculations of the mass-transfer
               process in two LMXB systems with 
               $P_{\rm orb}^{\rm ZAMS}=3.0$ days, $X=0.70, Z=0.02$ and
               $\alpha = 2.0$. 
               The mass of the donor stars are (a): $1.0\,M_{\odot}$,
               and (b): $1.6\,M_{\odot}$ -- top and bottom panels, respectively.
               See text for further details.
              }
  \end{figure*}
\addtocounter{figure}{-1}
  \begin{figure*}
    \psfig{file=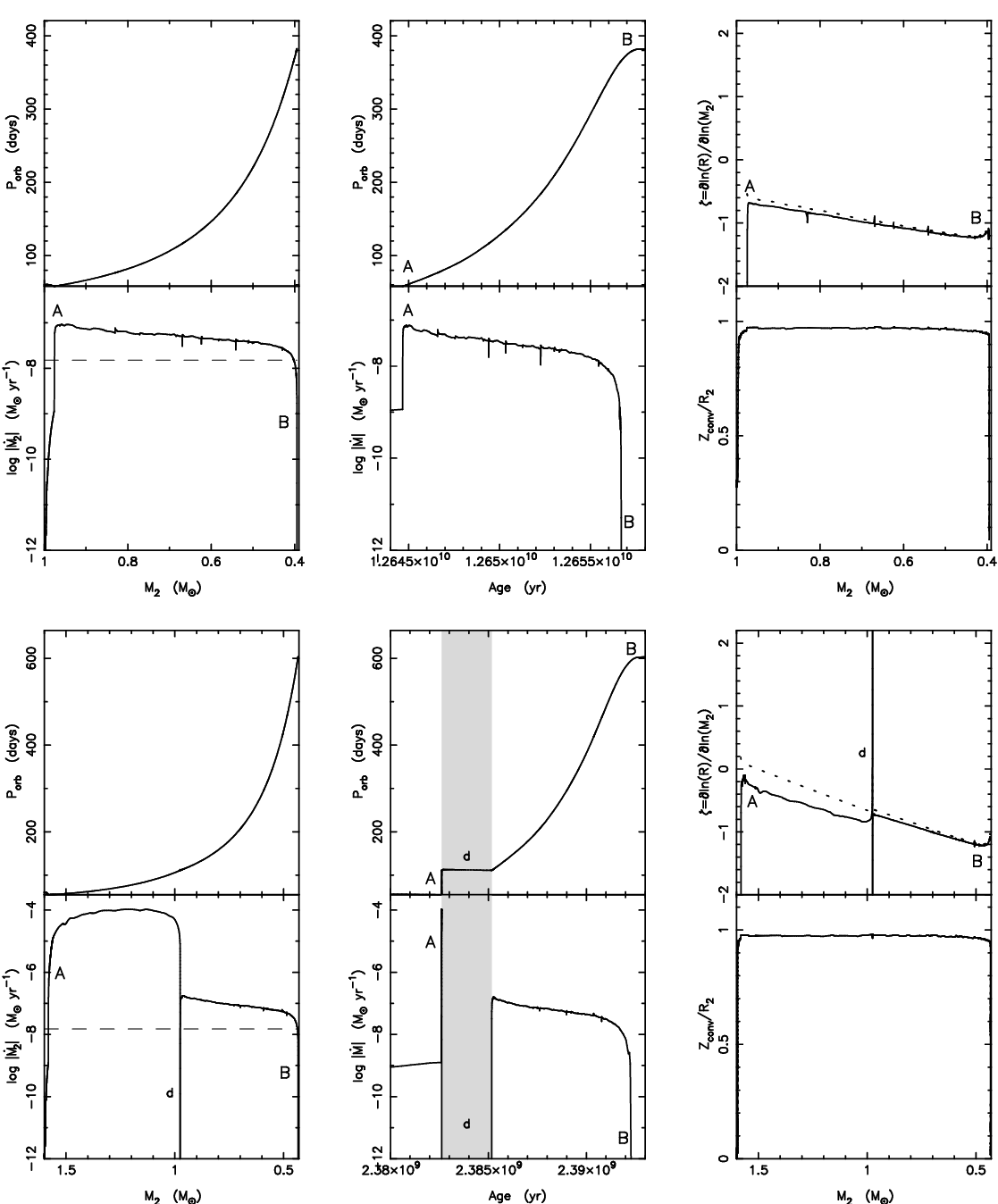,width=18.0cm,height=22.5cm}
    \caption[]{\hspace{-0.16cm}{\bf c-d.}\hspace{0.16cm}
               Numerical calculations of the mass-transfer
               process in two LMXB systems with 
               $P_{\rm orb}^{\rm ZAMS}=60.0$ days, $X=0.70, Z=0.02$ and
               $\alpha = 2.0$. 
               The mass of the doner stars are (c): $1.0\,M_{\odot}$,
               and (d): $1.6\,M_{\odot}$ -- top and bottom panels, respectively.
               See text for further details.
              }
  \end{figure*}

\subsection{Fig.~2a}
In Fig.~2a we adopted $M_2^{\rm i}=1.0\,M_{\odot}$ and
$P_{\rm orb}^{\rm ZAMS} = 3.0$ days.
In this case the time it takes for
the donor to become a (sub)giant and fill its Roche-lobe, to initiate
mass transfer, is 11.89 Gyr. Before the donor fills its Roche-lobe
the expansion due to shell hydrogen burning causes its moment of inertia
to increase which tends to slow down the rotation of the star.
However, the
tidal torques act to establish synchronization by transferring angular
momentum to the donor star at the expense of orbital angular momentum.
Hence at the onset of the mass transfer ($A$) the 
orbital period has decreased from the initial $P_{\rm orb}^{\rm ZAMS}=3.0$ 
days to $P_{\rm orb}^{\rm RLO}=1.0$ days and the
radius is now $R_2=R_{\rm L}=2.0\,R_{\odot}$.\\
We notice that the mass-loss rate of the donor star
remains sub-Eddington ($|\dot{M}_2| < \dot{M}_{\rm Edd}\approx
1.5\times 10^{-8}\,M_{\odot} \rm{yr}^{-1}$ for hydrogen-rich matter)
during the entire mass transfer\footnote{
Strictly speaking $\dot{M}_{\rm Edd}=\dot{M}_{\rm Edd}(R_{\rm NS})$
is slightly reduced during the accretion phase since the radius of
the neutron star decreases with increasing mass
({\em e.g.} for an ideal $n$-gas polytrope:
$R_{\rm NS}\propto M_{\rm NS}^{-1/3}$). However, this only
amounts to a correction of less than 20\% for various equations-of-state,
and thus we have not taken this effect into account.}. Thus we expect
all the transferred material to be accreted onto the neutron star, if disk
instabilities and propeller effects can be neglected (see Sections 5.7 and 6.4).
Therefore we have no mass loss from the system in this case 
-- i.e. $\dot{M}=0$. 
The duration of the mass-transfer phase for this system is quite long:
$\sim$1.0 Gyr ($11.89\rightarrow 12.91$ Gyr). 
At age, $t\sim 12.65$ Gyr ($P_{\rm orb}=5.1$ days; $M_2=0.317\,M_{\odot}$)
the donor star detaches slightly from its Roche-lobe ($d$)
and the mass transfer ceases temporarily for $\sim$ 25 Myr --
see next subsection for an explanation.\\
The Roche-radius exponent calculated from eq.~(17) is plotted as
a dotted line as a function of $M_2$ in the upper right panel.
However, our numerical calculations (full line) show that tidal effects 
are significant and increase $\zeta$ by $\sim\,$0.5--0.8 until 
$M_2 \approx 0.54\,M_{\odot}$ ($p$). At this point the magnetic braking
is assumed to switch off, since $Z_{\rm conv}/R_2 > 0.80$.
Note that during the mass transfer phase $\zeta \approx \zeta _{\rm L}$
and, as long as the mass transfer is not unstable on a dynamical
timescale, we typically have in our code: $1\times 10^{-4} <
\ln (R_2/R_{\rm L}) < 7\times 10^{-3}$ and hence practically
$\zeta = \zeta _{\rm L}$.\\
The final outcome for this system is a BMSP with an orbital
period of $P_{\rm orb}^{\rm f}=9.98$ days and a {He} white dwarf (WD)
with a mass of $M_{\rm WD}=0.245\,M_{\odot}$ ($B$). The final mass
of the NS is $M_{\rm NS}=2.06\,M_{\odot}$, since we
assumed all the material was accreted onto the NS given 
$|\dot{M}_2| < \dot{M}_{\rm Edd}$ during the entire
{X}-ray phase. However, in Section~6
we will discuss this assumption and the important question of disk
instabilities and the propeller mechanism in more detail. 

\subsection{Fig.~2b}
In Fig.~2b we adopted $M_2^{\rm i}=1.6\,M_{\odot}$ and
$P_{\rm orb}^{\rm ZAMS} = 3.0$ days. The RLO is initiated at an age of
$t=2.256$ Gyr when $P_{\rm orb}^{\rm RLO}=2.0$ days and $R_2=3.8\,R_{\odot}$
($A$).
In this case the mass-transfer rate is super-Eddington 
($|\dot{M}_2| > \dot{M}_{\rm Edd}$, cf. dashed line) at the beginning 
of the mass-transfer phase.
In our adopted model of "isotropic
re-emission" we assume all material in excess of the Eddington
accretion limit to be ejected from the system, while carrying with
it the specific orbital angular momentum of the neutron star.
Hence $|\dot{M}|=|\dot{M}_2|-\dot{M}_{\rm Edd}$.
Initially $|\dot{M}_2|\approx 10^2 \dot{M}_{\rm Edd}$ at the onset
of the RLO and then $|\dot{M}_2|$ decreases from 10 to $1\,\dot{M}_{\rm Edd}$
at $M_2\approx 0.7 M_{\odot}$ ($e$) before it becomes sub-Eddington for the
rest of the mass-transfer phase.
Mass loss from the system as a result of a Reimers wind in the red giant stage
prior to RLO (A) is seen to be less than $10^{-11}\,M_{\odot}$ yr$^{-1}$.
By comparing the different panels for the evolution, we notice that
the initial super-Eddington mass transfer phase ($A-e$) lasts for
22 Myr. In this interval the companion mass decreases from
$1.6\,M_{\odot}$ to $0.72\,M_{\odot}$. Then the system enters
a phase ($e-d$) of sub-Eddington mass transfer at $P_{\rm orb}$=5.31 days
which lasts for 41 Myr. When $M_2=0.458\,M_{\odot}$, and 
$P_{\rm orb}$=13.6 days, the system detaches and the X-ray source
is extinguished for about 40 Myr ($d$), cf. gray-shaded area. 
The temporary detachment is caused by a transient contraction of the donor star 
when its hydrogen shell source moves into the hydrogen rich layers left behind 
by the contracting convective core during the early main sequence stage. At the 
same time the convective envelope has penetrated inwards to its deepest point, 
i.e. almost, but not quite, to the {H}-shell source. 
The effect of a transient contraction of single low-mass stars evolving
up the RGB, as a result of a sudden discontinuity in the chemical
composition, has been known for many years 
(Thomas 1967; Kippenhahn \& Weigert 1990) but has hitherto escaped
attention in binary evolution.
After the transient 
contraction the star re-expands enough to fill its Roche-lobe again and further 
ascends the giant branch.
The corresponding final phase of mass transfer ($d-B$) 
is sub-Eddington ($|\dot{M}_2|\approx 0.2\,\dot{M}_{\rm Edd}$) and
lasts for 60 Myr. The end product of this binary is a recycled pulsar
and a {He}-WD companion with an orbital period of 41.8 days.
In this case we obtain $M_{\rm WD}=0.291\,M_{\odot}$ and 
$M_{\rm NS}=2.05\,M_{\odot}$.\\
The total duration of the mass-transfer phase during which the system is an
{\em active} {X}-ray source is
$t_{\rm X} = 123$ Myr (excluding the quiescence
phase of 40 Myr) which is substantially shorter
compared to the case discussed above (Fig.~2a).\\
The reason for the relatively wide final orbit of this system, compared
to the case discussed above with the $1.0\,M_{\odot}$ donor, is caused
by the super-Eddington mass transfer during which a total of
$0.55\,M_{\odot}$ is lost from the system.\\
The numerical calculations of $\zeta$ for this donor star (full line)
fits very well with our simple analytical expression (dotted line) which
indicates that the effects of the tidal spin-orbit interactions are not 
so significant in this case.
  \begin{figure}[t]
    \psfig{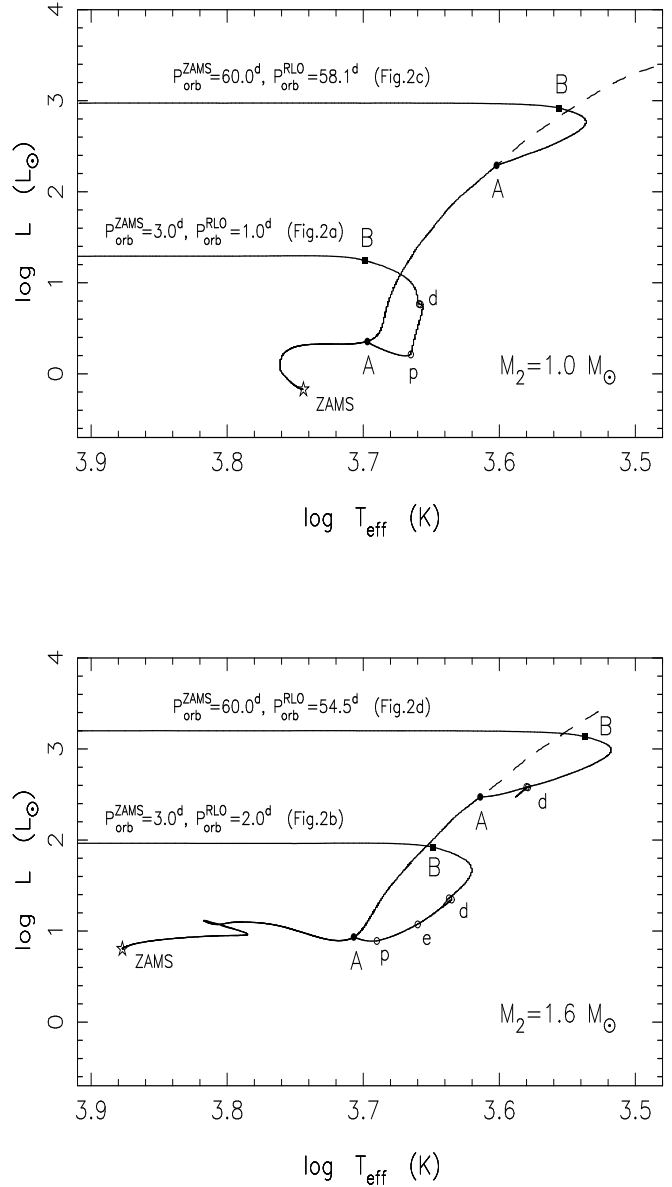}
     \caption{Calculated evolutionary tracks in the Hertzsprung-Russel 
              diagram of the four donor stars used in Fig.~2.
              The dashed lines represent the evolution of a single
              star with mass $M_2$. The mass-transfer phase (RLO) is
              initiated at $A$ and ceases at $B$.
              The symbols along the evolutionary tracks correspond
              to those indicated in Fig.~2.}
  \end{figure}

\subsection{Fig.~2c}
In this figure we adopted $M_2^{\rm i}=1.0\,M_{\odot}$ and
$P_{\rm orb}^{\rm ZAMS} = 60.0$ days. The RLO is initiated ($A$) at an age of
$t\simeq 12.645$ Gyr. At this stage the mass of the donor 
has decreased to $M_2^{\rm RLO}=0.976\,M_{\odot}$
as a result of the radiation-driven wind of the giant star.
However, the orbital period has also decreased
($P_{\rm orb}^{\rm RLO}=58.1$ days) and thus the shrinking
of the orbit due to tidal spin-orbit coupling
dominates over the widening of the orbit caused by the wind
mass loss.\\
The total interval of mass transfer is quite
short, $t_{\rm X}=13.3$ Myr.
The mass-transfer rate is super-Eddington during the entire
evolution ($|\dot{M}_2|\approx 1-6\,\dot{M}_{\rm Edd}$)
and therefore the NS only accretes very little material:
$\Delta M_{\rm NS} = \langle \dot{M}_{\rm Edd}\rangle\,\Delta t_{\rm mt} \approx
0.20\,M_{\odot}$. The reason for the high mass-loss rate of the donor star
is its deep convective envelope (see lower right panel). 
Since the initial configuration of
this system is a very wide orbit, the donor will be rather evolved 
on the RGB when it fills its Roche-lobe
($R_2=29.3\,R_{\odot}$ and $P_{\rm orb}^{\rm RLO}=58.1$ days). Hence the donor 
swells up in response to mass loss (i.e. $\zeta <0$) as a result of
the super-adiabatic temperature gradient in its giant envelope. 
The radius exponent is well described by our analytical formula
in this case. The final outcome of this system is a wide-orbit
($P_{\rm orb}^{\rm f}=382$ days) BMSP with a $\sim 0.40\,M_{\odot}$ {He}-WD
companion.

\subsection{Fig.~2d}
Here we adopted $M_2^{\rm i}=1.6\,M_{\odot}$ and
$P_{\rm orb}^{\rm ZAMS} = 60.0$ days. 
At the onset of the RLO the donor mass is $M_2^{\rm RLO}=1.582\,M_{\odot}$.
In this case we do not only have a giant donor with a deep convective
envelope. It is also (initially) heavier than the accreting NS.
Both of these circumstances makes it difficult for the donor
to retain itself inside its Roche-lobe once the mass transfer is
initiated. 
It is likely that such systems, with huge mass-transfer rates, 
evolve into a phase where matter piles up around the neutron star
and presumably forms a growing, bloated cloud engulfing it. 
The system could avoid a spiral-in when it manages to evaporate
the bulk of the transferred matter from the surface of the
(hot) accretion cloud via the liberated accretion energy.  
This scenario would require the radius of the accretion cloud, $r_{\rm cl}$
to be larger than $\sim R_{\rm NS}\,(|\dot{M}|/\dot{M}_{\rm Edd})$
in order for the liberated accretion energy to eject the transfered
material. However, if there is insufficient gas cooling $r_{\rm cl}$
could be smaller from an energetic point of view.
At the same time 
$r_{\rm cl}$ must be smaller than the Roche-lobe radius of the neutron star
(cf. eq.~10 with $q=M_{\rm NS}/M_2$) during the entire evolution.
In that case our simple isotropic re-emission model would approximately
remain valid. Assuming this to be the case we find the mass-transfer
rate is extremely high: $|\dot{M}_2|\approx 10^4\,\dot{M}_{\rm Edd}$
and more than $0.5\,M_{\odot}$ is lost from the donor (and the system)
in only a few $10^3$ yr. 
The system survives and the orbital
period increases from 54.5 days to 111 days during this short phase.\\
After this extremely short mass-transfer epoch, with an ultra-high
mass-transfer rate, the donor star relaxes ($d$) and shrink inside its 
Roche-lobe for 2.5 Myr
when $M_2=0.98\,M_{\odot}$. The mass transfer is resumed again
for 7.5 Myr at a more moderate super-Eddington rate ($d-B$).
The final outcome is a binary pulsar with a $0.43\,M_{\odot}$ {He}-WD
companion and an orbital period of 608 days. Though the NS only accretes
$\sim 0.10\,M_{\odot}$ as a result of the short integrated accretion
phase it will probably be spun-up sufficiently to become a millisecond
pulsar since millisecond pulsars evidently are also formed in systems 
which evolve e.g. through a CE with similar (or even shorter) phases
of accretion (van~den~Heuvel 1994b).\\
The initial extreme evolution of this system causes an offset in
$\zeta$ until the more moderate mass-transfer phase ($d-B$) continues
at $M_2=0.98\,M_{\odot}$. It should be noted that a system like
this is very unlikely to be observed in the ultra-high mass-transfer
state due to the very short interval ($<10^4$ yr) of this phase.

\subsection{$M_2 > 2\,M_{\odot}$, runaway mass transfer and onset of a CE}
The latter example above illustrates very well the situation near the
threshold for unstable mass transfer on a dynamical timescale and
the onset of a CE evolution\footnote{
We notice, that this very high mass-transfer rate
might lead to hyper-critical accretion onto the neutron star
and a possible collapse of the NS into a black hole
if the equation-of-state is soft (cf. Chevalier 1993; Brown \& Bethe 1994;
Brown 1995). However, new results obtained by Chevalier (1996) including
the centrifugal effects of a rotating infalling gas might change this 
conclusion.}.
If the donor star is heavier than $1.8\,M_{\odot}$ a critical overflow is likely
to occur since the orbit shrinks in response to mass transfer ($q<1.28$, cf.
Section~4). This is also the situation if $P_{\rm orb}$ is large because
the donor in that case develops a deep convective envelope which causes it
to expand in response to mass loss and a runaway mass transfer sets in.
When a runaway mass transfer sets in we were not able to prevent
it from critically overflowing its Roche-lobe and our code breaks down.
At this stage the neutron star
is eventually embedded in a CE with its companion and it will spiral
in toward the center of its companion as a result of removal of orbital
angular momentum by the drag force acting on it\footnote{
However, even binaries with donor stars of 2--6$\,M_{\odot}$
might survive the mass transfer avoiding a spiral-in phase in case
the envelope of the donor is still radiative at the onset of the RLO.}.
The final result 
of the CE depends mainly on the orbital period and the mass of the  
giant's envelope. If there is enough orbital energy available
(i.e. if $P_{\rm orb}$ is large enough at the onset of the CE), then
the entire envelope of the giant can be expelled as a result of the liberated 
orbital energy, which is converted into kinetic energy that provides
an outward motion of the envelope decoupling it from its core.
This leaves behind a tight binary with a heavy WD (the core of
the giant) and a moderately recycled pulsar. There are five such systems
observed in our Galaxy. They all have a {CO}-WD and $P_{\rm orb}=6\sim 8$ days.
These are the so-called class$\,${C} BMSPs.\\
If there is not enough orbital energy available to expel the envelope,
then the NS spirals in completely to the center of the giant 
and a Thorne-\.{Z}ytkow object is formed. Such an object might evolve
into a single millisecond pulsar (e.g. van~den~Heuvel 1994a) or
may collapse into a black hole (Chevalier 1996).

\subsection{The ($P_{\rm orb},M_{\rm WD}$) correlation}
We have derived new ($P_{\rm orb},M_{\rm WD}$) correlations based on
the outcome of the 121 LMXB models calculated for this work.
They are shown in Fig.~4a (the top panel). 
We considered models with donor star masses ($M_2$) in the interval 
$1.0-2.0\,M_{\odot}$, chemical compositions ranging from Population~{I}
({X}=0.70, {Z}=0.02) to Population~{II} ({X}=0.75, {Z}=0.001) and
convective mixing-length parameters $\alpha \equiv l/H_{\rm p}$ from 2--3
(here $l$ is the mixing length and $H_{\rm p}$ is the local pressure 
scaleheight).
Following Rappaport~et~al. (1995) we chose our standard model with
$M_2=1.0\,M_{\odot}$, Population~{I} composition and $\alpha =2$, cf.
thick line in Fig.~4. The upper limit of $M_2$ is set by the requirement that
the mass transfer in the binary must be dynamically stable, and the lower
limit by the requirement that the donor star must evolve off the main
sequence within an interval of time given by:
$t_{\rm ms} < t_{\rm Hubble} - t_{\rm gal} - t_{\rm cool}$.
Here $t_{\rm Hubble} \sim 15$ Gyr is the age of the Universe,
$t_{\rm gal}\sim 1$ Gyr is the minimum time between the Big~Bang and formation
of our Milky Way and $t_{\rm cool}\sim 3$ Gyr is a typical low 
value of WD companion cooling ages, following the mass-transfer phase,
as observed in BMSPs (Hansen \& Phinney 1998).
We thus find $M_2 \simeq 1.0\,M_{\odot}$ as a conservative lower limit.

The first thing to notice,
is that the correlation is more or less independent of the initial
donor star mass ($M_2$) -- only for $M_2\ga 2.0 M_{\odot}$ (where
the mass transfer becomes dynamically unstable anyway for
$P_{\rm orb}^{\rm i}\ga 4.2$ days) we see a slight deviation. 
This result is expected if $M_{2\,\rm core}$ (and therefore
$R_2$ and $P_{\rm orb}$) is independent of $M_2$.
We have performed a check on this statement using our calculations for
an evolved donor star on the RGB. As an example, in
Table~1 we have written $L$, $T_{\rm eff}$ and $M_{2\,\rm core}$
as a function of $M_2$ when it has
evolved to a radius of $50.0\,R_{\odot}$.
In addition we have written the mass of the donor's envelope
at the moment $R_2=50.0\,R_{\odot}$. 
\begin{table}[h]
\caption{Stellar parameters for a star with $R_2=50.0\,R_{\odot}$ -- see text.}
\begin{center}
\begin{tabular}{lllll}
\hline
\noalign{\smallskip}
$M_2/M_{\odot}$            &1.0$^{**}$ & 1.6$^{**}$ & 1.0$^{*}$ & 1.6$^{*}$ \\
\noalign{\smallskip}
\hline
\noalign{\smallskip}
$\log L/L_{\odot}$       & 2.566 & 2.624 & 2.644 & 2.723\\
$\log T_{\rm eff}$              & 3.554 & 3.569 & 3.573 & 3.593\\
$M_{\rm 2 core}/M_{\odot}$ & 0.336 & 0.345 & 0.342 & 0.354\\
$M_{\rm 2 env}/M_{\odot}$  & 0.215 & 0.514 & 0.615 & 1.217\\
\hline
\end{tabular}
\end{center}
  \begin{list}{}{}
    \item[*] Single star ({X}=0.70, {Z}=0.02 and $\alpha$=2.0). 
    \item[**] Binary donor ($P_{\rm orb}^{\rm ZAMS}=60.0$ days and 
                            $M_{\rm NS}=1.3\,M_{\odot}$)
\end{list}{}{}
\end{table}
We conclude, for a given 
chemical composition and mixing-length parameter, $M_{2\, \rm core}$
is practically independent of $M_2$ (to within a few per cent) and that
mass loss from the envelope via RLO has similar little influence on the
($R_2,M_{2\,\rm core}$) correlation as well. For other choices of
$R_2$ the differences were found to be smaller. In Fig.~5 we have shown 
$P_{\rm orb}$ (which resembles $R_2$) as a function of $M_{2\,\rm core}$.\\
Much more important is the theoretical uncertainty in the value of the
convective mixing-length parameter and most important is the initial
chemical composition of the donor star. 
We have estimated a ($P_{\rm orb},M_{\rm WD}$) correlation from an overall
best fit to all the models considered in Table~A1 and obtain 
($P_{\rm orb}$ in days):
\begin{equation}
  \frac{M_{\rm WD}}{M_{\odot}} = \left( \frac{P_{\rm orb}}{b} \right) ^{1/a}+\,c
\end{equation}
where, depending on the chemical composition of the donor,
\begin{equation}
(a,b,c) = \left\{ \begin{array}{llll}
       4.50\; & 1.2\times 10^5\; & 0.120\; & \mbox{\hspace{1.0cm}\rm{Pop.{I}}}\\
       4.75 & 1.1\times 10^5 & 0.115 & \mbox{\hspace{1.0cm}\rm{Pop.{I+II}}}\\
       5.00 & 1.0\times 10^5 & 0.110 & \mbox{\hspace{1.0cm}\rm{Pop.{II}}}
                 \end{array}
         \right.
\end{equation}
This formula is valid for binaries with: 
$0.18\la M_{\rm WD}^{\rm He}/M_{\odot} \la 0.45$.
The uncertainty in the initial chemical abundances of the donor star results
in a spread of a factor $\sim$1.4 about the 
median (Pop.{I+II}) value of $P_{\rm orb}$ at any given
value of $M_{\rm WD}$. The spread in the ($P_{\rm orb},M_{\rm WD}$) correlation 
arises solely from the spread in the ($R_2,M_{\rm 2\,core}$) correlation as
a result of the different chemical abundances, and/or $\alpha$, of
the giant donor star ascending the RGB.

  \begin{figure*}
    \psfig{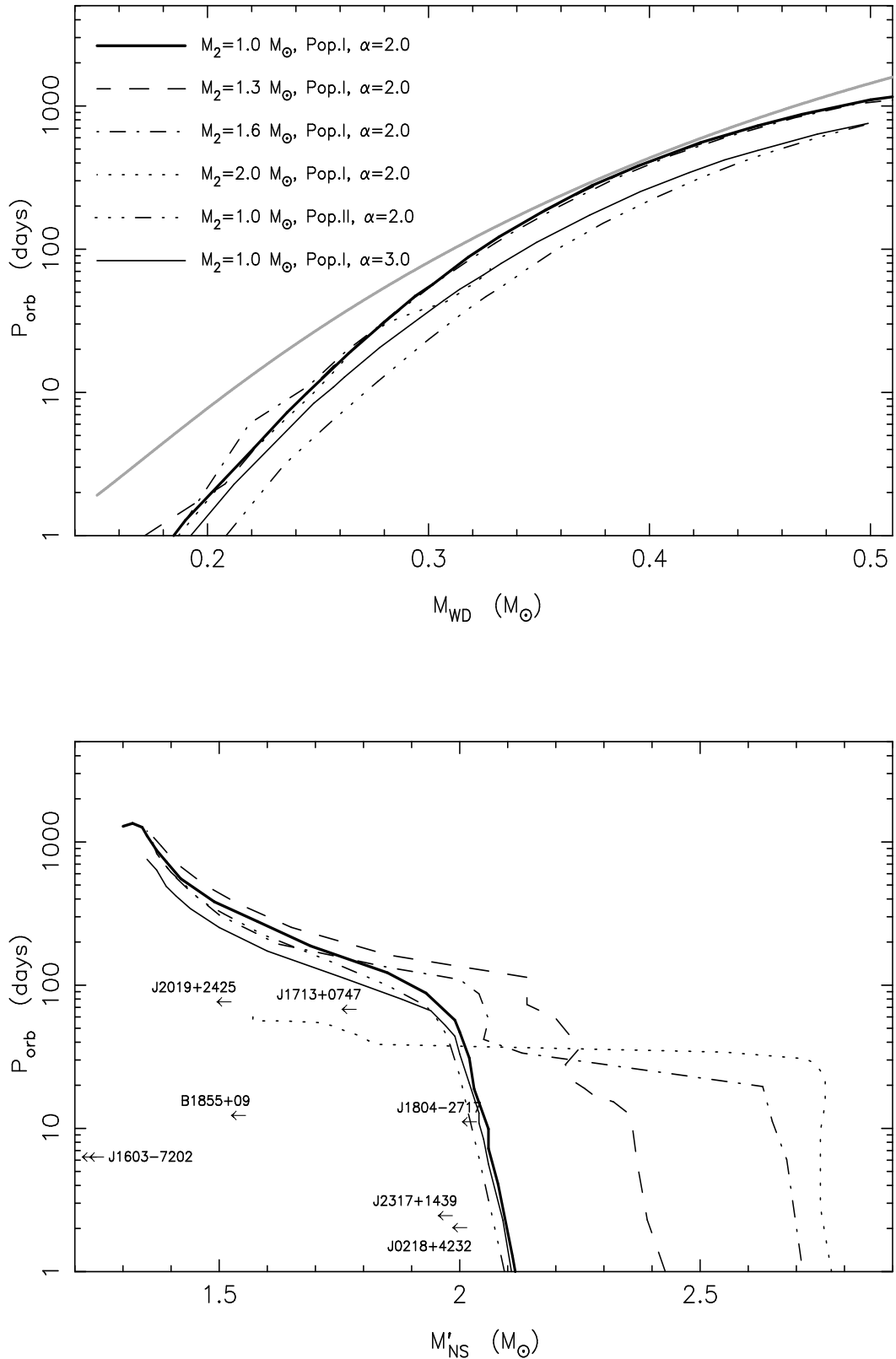}
    \caption[]{\hspace{-0.16cm}{\bf a-b.}\hspace{0.16cm}
               The ($P_{\rm orb},M_{\rm WD}$) correlation
               (top panel) and the ($P_{\rm orb},M_{\rm NS}$)
               anti-correlation (bottom panel) calculated
               for different donor star masses, $M_2$ chemical
               compositions (Pop.{I}: {X}=0.70; {Z}=0.02 and
               Pop.{II}: {X}=0.75; {Z}=0.001) and mixing-length 
               parameters, $\alpha$ as indicated in the top panel.
               The gray line shows the
               correlation obtained by Rappaport~et~al. (1995)
               for $M_2=1.0\,M_{\odot}$, Pop.{I} chemical abundances
               and $\alpha =2.0$.
               The post-accretion $M_{\rm NS}$ curves (bottom) 
               assume no mass loss from
               accretion disk instabilities of propeller effects
               -- see Sections 5.7 and 6.4.
              }
  \end{figure*}
If we compare our calculations with the work of Rappaport~et~al. (1995)
we find that our best fit results in significantly lower 
values of $P_{\rm orb}$ 
for a given mass of the WD in the interval $0.18\la M_{\rm WD}/M_{\odot}\la 
0.35$.
It is also notable that these authors find a maximum spread in $P_{\rm orb}$
of a factor $\sim$2.4 at fixed $M_{\rm WD}$. 
For $0.35\la M_{\rm WD}/M_{\odot}\la 0.45$ their results 
agree with our calculations to within 20$\,$\%.
A fit to our eq.~(20) with the results of Rappaport~et~al. (1995) 
yields: $a=5.75$, $b=8.42\times 10^4$ and $c=0$ (to an accuracy within 1$\,$\%
for $0.18 \la M_{\rm WD}/M_{\odot} \la 0.45$)
for their donor models with population$\,${I} chemical composition and $\alpha 
=2.0$.
For their Pop.$\,${II} donors we obtain $b=3.91\times 10^4$ and
same values for $a$ and $c$ as above.
We also obtain somewhat lower values of $P_{\rm orb}$, for a given
mass of the WD, compared with the results of Ergma, Sarna \& Antipova (1998).

The discrepancy between the results of the above mentioned papers
and our work is a result of different
input physics for the stellar evolution (cf. Section~2).
Ergma, Sarna \& Antipova (1998) uses models 
based on Paczynzki's code, and Rappaport~et~al. (1995) used an older version
of Eggleton's code than the one used for this work.
In our calculations we have also included the effects of tidal dissipation.
However, these effects can not account for the discrepancy 
since in this paper 
we only considered binaries with $P_{\rm orb}^{\rm i}>2$ days and thus
the effects of the tidal forces are relatively small 
(the contribution to the stellar
luminosity from dissipation of tidal energy is only
$L_{\rm tidal} \la 0.05 L_{\rm nuc}$ for $P_{\rm orb}=2$ days).
 
In analogy with Rappaport~et~al. (1995) and Ergma, Sarna \& Antipova (1998)
we find that, for a given value of
$M_{\rm WD}$, $P_{\rm orb}$ is decreasing with increasing $\alpha$, and
$P_{\rm orb}$ is increasing with increasing metallicity.
We find
$M_{\rm WD}^{z=0.001}\simeq M_{\rm WD}^{z=0.02}+0.03 M_{\odot}$ which
gives a stronger dependency on metallicity, by a factor $\sim$2, compared
to the work of Ergma, Sarna \& Antipova (1998).

It should be noticed, that the ($P_{\rm orb},M_{\rm WD}$) correlation 
is {\em independent} of $\beta$ (the fraction of transferred material lost from
the system), the mode of mass loss and degree of magnetic braking
since, as demonstrated above, the 
relationship between $R_2$ and $M_{\rm 2\,core}$ of the giant donors
remains unaffected by the exterior stellar conditions governing the
process of mass transfer 
-- see also Alberts~et~al. (1996).
But for the {\em individual} binary, $P_{\rm orb}$ and $M_{\rm WD}$
do depend on $\beta$ and they increase with increasing values of $\beta$
(see e.g. the bottom panel of Fig.~1 for $q<1$ which always applies near
the end of the mass-transfer phase). 
  \begin{figure}[t]
    \psfig{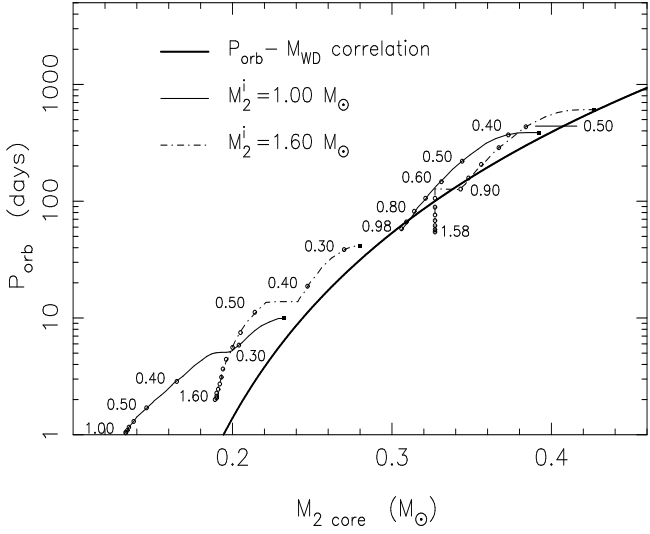}
     \caption{Evolutionary tracks of the four LMXBs in Figs~2 and 3,
              showing the binary orbital period changing as a function of
              the mass of the core of the donor star.
              At the termination of the mass-transfer process 
              $M_{\rm 2\,core}\approx M_{\rm WD}-0.01\,M_{\odot}$
              and the end-points of the evolutionary tracks
              are located near the
              curve representing the ($P_{\rm orb},M_{\rm WD}$)
              correlation. The initial orbital periods were
              $P_{\rm orb}^{\rm ZAMS}=3.0$ days and 
              $P_{\rm orb}^{\rm ZAMS}=60.0$ days for the two bottom
              and top tracks, respectively. Furthermore we used
              Population~{I} chemical abundances and $\alpha=2$.}
  \end{figure}

As mentioned in our examples earlier in this section, there is a competition
between the wind mass loss and the
tidal spin-orbit interactions for determining the orbital evolution
prior to the RLO-phase.
This is demonstrated in Fig.~6 where we have shown the changes
in $P_{\rm orb}$ and $M_2$, from the ZAMS stage to the onset of the RLO, 
as a function of the initial ZAMS orbital period.
It is seen that only for binaries with $P_{\rm orb}^{\rm ZAMS} > 100$ days
will the wind mass-loss be efficient enough to widen the orbit.
For shorter periods the effects of the spin-orbit interactions dominate
(caused by expansion of the donor) and loss of orbital angular
momentum causes the orbit to shrink.
  \begin{figure}[t]
    \psfig{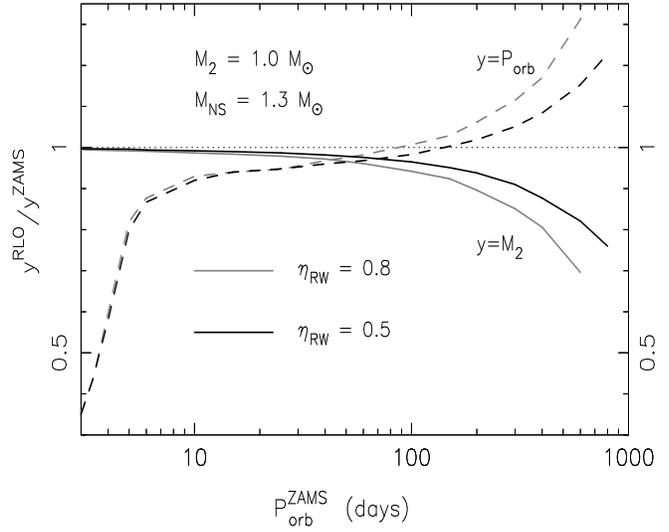}
     \caption{The changes of donor mass, $M_2$ (full lines) and orbital
              period, $P_{\rm orb}$ (dashed lines), due to wind mass loss
              and tidal spin-orbit interactions, from the ZAMS
              until the onset of the RLO as a function of the
              initial orbital period. Plots are shown for two different
              values of the Reimers' mass-loss parameter, $\eta_{\rm RW}$. 
              The binary is assumed to be circular.
              See text for further discussion.}
  \end{figure}

\subsection{The ($P_{\rm orb},M_{\rm NS}$) anti-correlation}
We now investigate the interesting relationship between the final mass
of the NS and the final orbital period. In Fig.~4b (the bottom panel)
we have plotted $P_{\rm orb}$
as a function of the potential maximum mass of the recycled pulsar,
$M_{\rm NS}^{\prime}$. This value is the final mass of the NS if
mass loss resulting from instabilities in the accretion process are
neglected. Another (smaller) effect which has also been ignored is the
mass deficit of the accreted material as it falls deep down 
the gravitational potential well of the NS.
The gravitational mass of a NS (as measured from a distant observer
by its gravitational effects) contains not only the rest mass of the
baryons, but also the mass equivalent of the negative binding energy,
$\Delta M_{\rm def}=E_{\rm bind}/c^2<0$. Depending on the equation-of-state
$\Delta M_{\rm def}\sim$10\% of the gravitational mass
(Shapiro \& Teukolsky 1983). This is hence also the efficiency of radiative
emission in units of available rest-mass energy incident on the NS. 
Thus we can express the actual post-accretion gravitational mass of a
recycled pulsar by ($\partial m_2<0$):
\begin{equation}
  M_{\rm NS} = M_{\rm NS}^{\rm i} + \left[-\int _{M_2}^{M_{\rm WD}}
                (1-\beta ')\,\partial m_2 
                - \Delta M_{\rm dp}\right] \,k_{\rm def}
\end{equation}
Here $\beta '\equiv\max\left((|\dot{M}_2|-\dot{M}_{\rm Edd})/|\dot{M}_2|, 
0\right)$
is the fraction of material lost in a relativistic jet as a result of 
super-Eddington mass transfer;
$\Delta M_{\rm dp}=\Delta M_{\rm disk}+ \Delta M_{\rm prop}$ 
is the sum of matter lost from the accretion disk (as a result
of viscous instabilities or wind corona) and matter being 
ejected near the pulsar
magnetosphere as a result of the centrifugal propeller effect, and finally 
$k_{\rm def}=\langle \frac{M_{\rm NS}}{M_{\rm NS}-\Delta M_{\rm def}} \rangle
 \approx 0.90$ is a factor that expresses the ratio of gravitational mass
to rest mass of the material accreted onto the NS.
$M_{\rm NS}^{\prime}$ used in Fig.~4b is given by the expression above
assuming $\Delta M_{\rm dp}=0$ and $\Delta M_{\rm def}=0$ ($k_{\rm def}=1$).

We see that the ($P_{\rm orb},M_{\rm NS}^{\prime}$) anti-correlation 
is more or less
independent of the chemical composition and $\alpha$ of the donor star,
whereas it depends strongly on $M_2$ for $P_{\rm orb} \la 50$ days.
This anti-correlation between $P_{\rm orb}$ and
$M_{\rm NS}^{\prime}$ is quite easy to understand: binaries with large initial
orbital periods will have giant donor stars with deep convective
envelopes at the onset of the mass transfer; hence the mass-transfer rate
will be super-Eddington and subsequently a large fraction of the
transferred material will be lost from the system. Therefore BMSPs
with large values of $P_{\rm orb}$ are expected to have relatively
light NS  -- cf. Sections 5.3 (Fig.~2c) and 5.4 (Fig.~2d).
Similarly, binaries with small values of $P_{\rm orb}^{\rm ZAMS}$ will
result in BMSPs with relatively small $P_{\rm orb}$ and large values of
$M_{\rm NS}^{\prime}$, since $\dot{M}_2$ will be sub-Eddington and
thus the NS has the potential to accrete all of the transferred material
-- cf. Sections 5.1 (Fig.~2a) and 5.2 (Fig.~2b). Therefore, if disk 
instabilities, wind corona and propeller effects were 
unimportant we would expect to find an
($P_{\rm orb},M_{\rm NS}$) anti-correlation among the observed BMSPs.
However, below (Section~6.2) we demonstrate that mass ejection arising from 
these effects is indeed important and 
thus it is very doubtful whether an ($P_{\rm orb},M_{\rm NS}$)
anti-correlation will be found from future observations.

\section{Discussion and comparison with observations}
\subsection{Comparison with condensed polytrope donor models}
Hjellming \& Webbink (1987) studied the adiabatic properties
of three simple families of polytropes by integrating the non-linear
Lane-Emden equation in Lagrangian coordinates.
The condensed polytropes, consisting of $n=3/2, \gamma =5/3$ (convective) 
envelopes with {He}-core point masses, are suitable for red giant stars.
In is not trivial to directly compare our calculations with e.g. the 
stability analysis of Soberman, Phinney \& van~den~Heuvel (1997) and
Kalogera \& Webbink (1996) since the donor does not 
restore (thermal) equilibrium after initiation of an unstable
mass-transfer process.
But it is important to point out, that systems which initiate RLO
with thermally unstable mass transfer could, in some cases, survive
this temporary phase -- even if $|\dot{M}_2|$ exceeds the Eddington
accretion limit by as much as a factor $\sim$10$^4$ (see Fig.~2d).
Similarly, systems which begin mass transfer on a thermal timescale
may in some cases (if $M_2$ is large compared to $M_{\rm NS}$)
eventually become dynamically unstable.
These results were also found by Hjellming (1989) and Kalogera \& Webbink (1996),
and we refer to these papers for a more detailed discussion on the
fate of thermally unstable systems.
Therefore it is not always easy to predict the final outcome
of an LMXB system given its initial parameters --
especially since the onset criteria of a CE phase is rather uncertain.
Nevertheless, we can conclude that LMXBs with $M_2 \le 1.8\,M_{\odot}$
will always survive the mass transfer phase. Systems with donor stars
$M_2 \ge 2\,M_{\odot}$ only survive if $P_{\rm orb}^{\rm ZAMS}$ is
within a certain interval.

Soberman, Phinney \& van~den~Heuvel (1997) also used the polytropic
models of Hjellming \& Webbink (1987) to follow the mass transfer
in binaries. The global outcome of such calculations are
reasonably good. However, 
the weakness of the polytropic models is that whereas they yield the
radius-exponent at the onset of the mass transfer, and the
approximated stellar structure at that given moment, they do not
trace the response of the donor very well during the mass-transfer phase.
The structural changes of the donor star
(e.g. the outward moving H-shell and the inward moving convection zone
giving rise to the transient detachment of the donor from its Roche-lobe)
can only be followed in detail by a Henyey-type iteration scheme
for a full stellar evolutionary model.

\subsection{The observed ($P_{\rm orb},M_{\rm WD}$) correlation}
The companion mass, $M_{\rm WD}$ of an observed binary pulsar is constrained
from its Keplerian mass function which is obtained from the
observables $P_{\rm orb}$ and $a_{\rm p} \sin i$:
\begin{equation}
  f\,(M_{\rm NS},M_{\rm WD}) = \frac {(M_{\rm WD} \,\sin i)^{3}}
                             {(M_{\rm NS} + M_{\rm WD})^{2}}
                        = \frac {4 \pi ^{2}}{G} \:
                          \frac {(a_{\rm p} \sin i)^{3}}{P_{\rm orb}^{2}}
\end{equation}
Here $i$ is the inclination angle (between the orbital angular momentum
vector and the line-of-sight toward the observer) and
$a_{\rm p}=a\,(M_{\rm WD}/(M_{\rm NS}+M_{\rm WD}))$
is the semi-major axis of the pulsar in a c.m. reference frame.
The probability of observing a binary system at an inclination angle
$i$, less than some value $i_0$, is $P(i<i_0)=1-\cos\,(i_0)$.\\
As mentioned earlier, there are indeed problems
with fitting the observed low-mass binary pulsars onto a
theoretical core-mass period relation.
The problem is particularly pronounced for the very wide-orbit BMSPs. Although
the estimated masses of the companions are quite uncertain (because of the
unknown orbital inclination angles and $M_{\rm NS}$) 
no clear observed ($P_{\rm orb},M_{\rm WD}$) correlation seems to be
present -- opposite to what is proposed by
several authors (e.g. Phinney \& Kulkarni 1994, Lorimer~et~al. 1996 and
Rappaport~et~al. 1995). In Table~A2 in the Appendix 
we have listed all the observed galactic (NS+WD) binary pulsars
and their relevant parameters. 
It was noticed by Tauris (1996) that the five BMSPs with $P_{\rm orb} > 100$
days all seem to have an observed $M_{\rm WD}^{\rm obs}$ which is lighter
than expected from the theoretical correlation
(at the $\sim$80$\,$\% confidence level on average).
There does not seem to be any observational selection effects which can
account for this discrepancy (Tauris 1996; 1998) -- {\em i.e.} why we 
should preferentially observe systems with small inclination angles
(systematic small values of $i$, rather than a random distribution,
would increase $M_{\rm WD}^{\rm obs}$ for the given observed mass functions
and thus the observations would match the theory).
Evaporation of the companion star, from a wind of relativistic
particles after the pulsar turns on, also seems unlikely since the
evaporation timescale (proportional to $P_{\rm orb}^{4/3}$)
becomes larger than $t_{\rm Hubble}$ for such wide orbits.
It is also worth mentioning that the orbital period change due to
evaporation, or general mass-loss in the form of a stellar wind,
is at most a factor of $\sim$2, if one assumes the specific angular
momentum of the lost matter is equal to that of the donor star.\\
Beware that the ($P_{\rm orb},M_{\rm WD}$) correlation is {\em not}
valid for BMSPs with {CO}/{O-Ne-Mg} WD companions as these systems did not 
evolve through a phase with stable mass transfer. The exception here
are the very wide orbit systems with $P_{\rm orb}^{\rm f}\ga 800$ days.
PSR B0820+02 might be an example of such a system. From Table~A1 (Appendix)
it is seen that we expect a maximum orbital period of $\sim\!1400$ days
for the NS+WD binaries. Larger periods are, of course, possible but
the binaries are then too wide for the neutron star to be recycled
via accretion of matter.

It should also be mentioned that the recycling 
process is expected to align the spin axis of the neutron star with 
the orbital angular momentum vector as a result of $>\!10^7$ yr of stable
disk accretion. Hence we expect (Tauris 1998)
the orbital inclination angle, $i$ to be equivalent to (on average)
the magnetic inclination angle,~$\alpha _{\rm mag}$ defined as the
angle between the pulsar spin axis and the center of the pulsar beam
({\em viz.} line-of-sight to observer).

\subsection{PSR J2019+2425}
PSR J2019+2425 is a BMSP with $P_{\rm orb}=76.5$ days and a
mass function $f=0.0107 M_{\odot}$ (Nice, Taylor \& Fruchter 1993).
In a recent paper (Tauris 1998) it was demonstrated that for this
pulsar $M_{\rm NS}\simeq 1.20\,M_{\odot}$, if the ($P_{\rm orb},M_{\rm WD}$)
correlation obtained by Rappaport~et~al. (1995) was taken at face value.
This value of $M_{\rm NS}$ is significantly lower than that of any other
estimated pulsar mass (Thorsett \& Chakrabarty 1999). 
However with the new ($P_{\rm orb},M_{\rm WD}$) correlation presented
in this paper we obtain a larger maximum mass ($i=90\degr$) of this
pulsar: $M_{\rm NS}^{\rm max}=1.39\,M_{\odot}$ or $1.64\,M_{\odot}$ for
a donor star of Pop.{I} or Pop.{II} chemical composition, respectively.
This result brings the mass of PSR~J2019+2425 inside the interval
of typical estimated values of $M_{\rm NS}$.

\subsection{$M_{\rm NS}$: dependence on the propeller effect 
            and accretion disk instabilities}
It is still an open question whether or not a significant amount
of mass can be ejected from an accretion disk as a result of
the effects of disk instabilities (Pringle 1981; van Paradijs 1996). 
However, there is clear evidence from observations of Be/{X}-ray transients
that a strong braking torque acts on these neutron stars which 
spin near their equilibrium periods. The hindering of accretion onto these
neutron stars is thought to be caused by their strong rotating magnetic
fields which eject the incoming material via centrifugal acceleration --
the so-called propeller effect (Illarionov \& Sunyaev 1985).

For a given observed BMSP we know $P_{\rm orb}$ and using
eqs~(20), (21) we can find $M_{\rm WD}$ for an adopted chemical
composition of the donor star. Hence we are also able to
calculate the maximum gravitational mass of the pulsar, $M_{\rm NS}^{\rm max}$
(which is found for $i=90\degr$, cf. eq.~23) 
since we know the mass function, $f$ from observations.
This semi-observational constraint on $M_{\rm NS}^{\rm max}$ can then be
compared with our calculations of $M_{\rm NS}^{\prime}$ (cf. Section~5.7).
The interesting cases are those where $M_{\rm NS}^{\prime}>M_{\rm NS}^{\rm max}$
(after correcting $M_{\rm NS}^{\prime}$ for the mass deficit).
These systems {\em must} therefore have lost matter ($\Delta M_{\rm dp}\neq 0$),
from the accretion disk or as a result of the propeller effect, in addition
to what is ejected when $|\dot{M}_2| > \dot{M}_{\rm Edd}$. These binaries are
plotted in Fig.~4b assuming an `intermediate' Pop.$\,${I}+{II}
chemical composition for the progenitor of the white dwarf. 
We notice that in a some cases 
we must require $\Delta M_{\rm dp}\simeq 0.50\,M_{\odot}$, or even more
for $M_2 > 1.0\,M_{\odot}$, in order to get $M_{\rm NS}$ below
the maximum limit ($M_{\rm NS}^{\rm max}$) indicated by the plotted arrow. 
We therefore conclude that mass ejection, in addition to what is caused by
super-Eddington mass-transfer rates, is very important in LMXBs.
Whether or not this conclusion
is equally valid for super- and sub-Eddington accreting systems is
difficult to answer since systems which evolve through an {X}-ray phase 
with super-Eddington mass-transfer rates lose a large amount of matter from the
system anyway and therefore naturally end up with small values of 
$M_{\rm NS}^{\prime}$.

\subsection{Kaon condensation and the maximum mass of NS}
It has recently been demonstrated (Brown \& Bethe 1994; Bethe \& Brown 1995)
that the introduction of kaon condensation sufficiently softens the
equation-of-state of dense matter, so that NS with masses more than
$\sim 1.56\,M_{\odot}$ will not be stable and collapse into a black hole.
If this scenario is correct, then we expect a substantial fraction
of LMXBs to evolve into black hole binaries 
-- unless $\Delta M_{\rm dp}$ is comparable to the difference between
$M_2$ and $M_{\rm WD}$ as indicated above.
However, it has recently been reported by
Barziv~et~al. (1999) that the HMXB Vela {X}-1 has a minimum value for the
mass of the neutron star of $M_{\rm NS}>1.68\,M_{\odot}$ at the 99$\,$\%
confidence level. It is therefore still uncertain at what critical mass
the NS is expected to collapse into a black hole.

\subsection{PSR J1603--7202}
The maximum allowed value of the pulsar mass in this system
is extremely low compared to other BMSP systems
with {He}-WD companions. We find $M_{\rm NS}^{\rm max}=0.96-1.11\,M_{\odot}$ 
depending on the chemical abundances of the white dwarf progenitor.
It is therefore quite suggestive that this system did not evolve
like the other BMSPs with a {He}-WD companion. 
Furthermore (as noted by Lorimer~et~al. 1996), it has
a relatively slow spin period of $P_{\rm spin}=14.8$ ms and $P_{\rm orb}=6.3$
days. Also its location in the ($P$,$\dot{P}$) diagram is atypical for
a BMSP with a {He}-WD (Arzoumanian, Cordes \& Wasserman 1999). 
All these characteristica are in common with BMSPs which possibly
evolved through a CE
evolution (van~den~Heuvel 1994b; Camilo 1996). We conclude therefore,
that this system evolved through a phase with critical unstable 
mass-transfer (like in a CE) and hence most likely hosts a {CO}-WD companion
rather than a {He}-WD companion. The latter depends on whether or not
helium core burning was ignited, and thus on the value of
$P_{\rm orb}^{\rm i}$ and $M_2$.
Spectroscopic observations should answer this question. 

\section{Conclusions}
\begin{itemize}
\item We have adapted a numerical computer code, based on Eggleton's
      code for stellar evolution, in order to carefully study the details
      of mass-transfer in LMXB systems. 
      We have included, for the first time to our knowledge,
      other tidal spin-orbit couplings than magnetic braking and also
      considered wind mass-loss during the red giant stage of the donor star.
\item We have re-calculated the ($P_{\rm orb},M_{\rm WD}$) correlation for
      binary radio pulsar systems using new input physics of stellar evolution
      in combination with detailed binary interactions.
      We find a correlation which yields a larger value of
      $M_{\rm WD}$ for a given value of $P_{\rm orb}$ compared to previous
      work.
\item Comparison between observations of BMSPs and our
      calculated post-accretion $M_{\rm NS}$ suggests that a large
      amount of matter is lost from the LMXBs; probably as a result of either
      accretion disk instabilities or the propeller effect.
      Hence it is doubtful whether or not observations will reveal
      an ($P_{\rm orb},M_{\rm NS}$) anti-correlation which would
      otherwise be expected from our calculations.
\item The mass-transfer rate is a strongly increasing function of
      initial orbital period and donor star mass.
      For relatively close systems with light donors ($P_{\rm orb}^{\rm 
ZAMS}<10$ 
days
      and $M_2 <1.3\,M_{\odot}$) the mass-transfer rate is sub-Eddington,
      whereas it can be highly super-Eddington by a factor of $\sim\!10^4$ for
      wide systems with relatively heavy donor stars ($1.6\sim 2.0\,M_{\odot}$),
      as a result of their deep convective envelopes. Binaries with
      (sub)giant donor stars with mass in excess of $\sim\!2.0\,M_{\odot}$ are 
      unstable to dynamical timescale mass loss. Such systems will evolve 
through
      a common envelope evolution leading to a short ($< 10$ days) orbital
      period BMSP with a heavy {CO}/{O-Ne-Mg} white dwarf companion.
      Binaries with unevolved heavy ($>2\,M_{\odot}$) donor stars might be
      dynamically stable against a CE, but also end up with a relatively
      short $P_{\rm orb}$ and a {CO}/{O-Ne-Mg} WD.
\item Based on our calculations, we present new evidence that
      PSR J1603--7202 did not evolve through a phase with stable mass transfer
      and that it is most
      likely to have a {CO} white dwarf companion.
\item The pulsar mass of PSR J2019+2425 now fits within the standard
      range of measured values for $M_{\rm NS}$, given our new
      ($P_{\rm orb},M_{\rm WD}$) correlation. 
\end{itemize}

\acknowledgements{We would like to thank Ed van~den~Heuvel for several 
discussions on many issues; Guillaume Dubus for discussions
on accretion disk instabilities; J{\o}rgen Christensen-Dalsgaard for
pointing out the well-known tiny loop in the evolutionary tracks of low-mass
stars on the RGB and Lev Yungelson for comments on the manuscript.
T.M.T. acknowledges the receipt of a Marie Curie Research Grant
from the European Commission.}

\appendix
\section{Tidal torque and dissipation rate}
We estimate the tidal torque due to the interaction between the tidally induced
flow and the convective motions in the stellar envelope by means of the simple
mixing-length model for turbulent viscosity $\nu=\alpha H_{\rm p} V_{\rm c}$, where the
mixing-length parameter $\alpha$ is adopted to be 2 or 3, $H_{\rm p}$ is the local
pressure scaleheight, and $V_{\rm c}$ the local characteristic convective velocity.
The rate of tidal energy dissipation can be expressed as (Terquem~et~al. 1998):
\begin{equation}
   \drv{E}{t}=-\frac{192 \pi}{5} \Omega^2 \int_{R_i}^{R_o} \rho r^2
  \nu \left[\left(\pdrv{\xi_r}{r}\right)^2+6 \left(\pdrv{\xi_h}{r}\right)^2
  \right] \, dr 
\end{equation}
where the integration is over the convective envelope and $\Omega$ is the
orbital angular velocity, i.e.  we neglect effects of stellar rotation. The
radial and horizontal tidal displacements are approximated here by
the values for the adiabatic equilibrium tide:  
\begin{equation}
  \xi_r= f r^2 \rho \left(\drv{P}{r}\right)^{-1}
\end{equation}
\begin{equation}
  \xi_h=\frac{1}{6 r} \drv{(r^2 \xi_r)}{r} 
\end{equation}
where for the dominant quadrupole tide ($l\!=\!m\!=2$) $f=\frac{-GM_2}{4a^3}$.\\
The locally dissipated tidal energy is taken into account as an extra energy
source in the standard energy balance equation of the star, while the
corresponding tidal torque follows as:
\begin{equation}
  \Gamma = - \frac{1}{\Omega} \drv{E}{t} 
\end{equation}
The thus calculated tidal angular momentum exchange $dJ= \Gamma dt$ between the
donor star and the orbit during an evolutionary timestep dt is taken into
account in the angular momentum balance of the system. If the so calculated
angular momentum exchange is larger than the amount required to keep the donor
star synchronous with the orbital motion of the compact star we adopt
a smaller tidal angular momentum exchange (and corresponding tidal dissipation
rate in the donor star) that keeps the donor star exactly synchronous.

\setlength{\tabcolsep}{2.6pt}
\begin{table*}
\caption{LMXB systems calculated for the work presented in this paper.
Pop.$\,${I} and Pop.$\,${II} chemical compositions correspond to
{X}=0.70, {Z}=0.02 and {X}=0.75, {Z}=0.001, respectively. $\alpha$ is the
mixing-length parameter. $P_{\rm orb}$ is in units of days and the
masses are in units of $M_{\odot}$.}
\begin{center}
\begin{tabular}{ccrccrcccrccrcccrccr}
\cline{1-6}\cline{8-13}\cline{15-20}
\noalign{\smallskip}
\cline{1-6}\cline{8-13}\cline{15-20}
\noalign{\smallskip}
      \multicolumn{7}{l}
        {\qquad $M_2=1.0 \qquad \alpha =2.0 \qquad$ Pop.$\,${I}} &
      \multicolumn{7}{l}
        {\qquad $M_2=1.0 \qquad \alpha =2.0 \qquad$ Pop.$\,${II}} &
      \multicolumn{6}{l}
        {\qquad $M_2=1.0 \qquad \alpha =3.0 \qquad$ Pop.$\,${I}}\\
\noalign{\smallskip}
\cline{1-6}\cline{8-13}\cline{15-20}
\noalign{\smallskip}
 $P_{\rm orb}^{\rm ZAMS}$ & $P_{\rm orb}^{\rm RLO}$ & $P_{\rm orb}^{\rm f}$ &
 $M_{\rm WD}^{\rm theo}$ & $M_{\rm NS}^{\prime}$ & $t_{\rm X}$ & &
 $P_{\rm orb}^{\rm ZAMS}$ & $P_{\rm orb}^{\rm RLO}$ & $P_{\rm orb}^{\rm f}$ &
 $M_{\rm WD}^{\rm theo}$ & $M_{\rm NS}^{\prime}$ & $t_{\rm X}$ & &
 $P_{\rm orb}^{\rm ZAMS}$ & $P_{\rm orb}^{\rm RLO}$ & $P_{\rm orb}^{\rm f}$ &
 $M_{\rm WD}^{\rm theo}$ & $M_{\rm NS}^{\prime}$ & $t_{\rm X}$\\
\noalign{\smallskip}
\cline{1-6}\cline{8-13}\cline{15-20}
\noalign{\smallskip}
  2.6&0.64& 0.09& 0.133&2.17& -- &&  3.0&0.73& 0.56&0.195&2.11&3090&&  2.5&0.59& 0.36&0.168&2.13&4400\\
  2.7&0.72& 1.28& 0.190&2.11&2790&&  3.1&0.82& 3.16&0.235&2.06&1570&&  2.6&0.68& 2.28&0.212&2.09&1980\\
  2.8&0.81& 4.08& 0.221&2.08&1740&&  3.2&0.94& 6.25&0.256&2.04&1110&&  2.7&0.80& 5.63&0.237&2.06&1180\\
  2.9&0.94& 7.24& 0.236&2.06&1240&&  3.4&1.26& 11.9&0.277&2.02& 770&&  2.8&0.94& 8.38&0.248&2.05& 930\\
  3.0&1.04& 9.98& 0.245&2.06& 990&&  4.0&2.12& 23.4&0.300&2.00& 430&&  2.9&1.06& 10.9&0.257&2.04& 730\\
  3.4&1.48& 18.8& 0.264&2.03& 613&&  5.0& 4.0& 38.3&0.318&1.98& 220&&  3.0&1.17& 12.8&0.262&2.04& 645\\
  4.0&2.32& 30.9& 0.280&2.02& 350&&  6.0& 5.2& 46.6&0.326&1.97& 174&&  3.4&1.60& 20.6&0.278&2.02& 400\\
  5.0& 4.0& 47.1& 0.294&2.00& 190&&  8.0& 7.3& 59.9&0.337&1.95& 140&&  4.0&2.87& 33.0&0.296&2.00& 212\\
  6.0& 5.2& 57.0& 0.302&1.99& 145&& 10.0& 9.2& 71.9&0.344&1.93&85.0&&  5.0& 4.2& 44.0&0.307&1.99& 138\\
 10.0& 9.2& 88.0& 0.318&1.93&72.0&& 15.0&14.1& 99.5&0.358&1.85&54.0&&  6.0& 5.3& 52.3&0.314&1.97& 106\\
 15.0&14.1&121.9& 0.332&1.85&46.3&& 25.0&23.8&151.8&0.379&1.72&32.3&&  8.0& 7.3& 66.3&0.325&1.94&68.4\\
 25.0&23.7&187.6& 0.353&1.69&28.5&& 40.0&38.5&226.3&0.402&1.59&21.1&& 10.0& 9.2& 79.7&0.333&1.88&51.5\\
 40.0&38.4&277.9& 0.374&1.58&18.5&& 60.0&58.5&314.2&0.423&1.51&14.7&& 15.0&14.1&111.7&0.349&1.76&34.2\\
 60.0&58.1&381.9& 0.395&1.49&13.3&& 80.0&78.7&392.3&0.439&1.46&11.6&& 25.0&23.7&173.0&0.373&1.60&21.0\\
  100&98.3&554.4& 0.423&1.42& 8.6&&  100&99.3&462.3&0.452&1.44& 9.5&& 40.0&38.4&252.5&0.396&1.50&13.6\\
  150& 150&727.8& 0.449&1.39& 6.2&&  150& 152&613.0&0.478$^{*}$&1.40& 6.7&& 60.0&58.2&342.7&0.418&1.44& 9.5\\
  200& 204&873.4& 0.469$^{*}$&1.37& 6.1&&  200& 207&740.9&0.498$^{*}$&1.38& 5.3&& 80.0&78.6&420.0&0.434&1.41& 7.5\\
\cline{8-13}
  300& 315&1104.0&0.500$^{*}$&1.35& 3.5&&     &    &     &     &    &    &&  100&99.0&489.2&0.449&1.39& 6.3\\
  400& 433&1266.5&0.528$^{*}$&1.34& 2.4&&     &    &     &     &    &    &&  150& 151&635.7&0.476$^{*}$&1.37& 4.4\\
  600& 692&1349.0&0.596$^{*}$&1.32& 1.1&&     &    &     &     &    &    &&  200& 206&756.4&0.499$^{*}$&1.35& 3.4\\
\cline{15-20}
  800& 982&1285.6&0.668$^{*}$&1.30& 0.4&&     &    &     &     &    &    &&     &    &     &     &    &    \\
\cline{1-6}
\noalign{\bigskip}
\noalign{\bigskip}
\cline{1-6}\cline{8-13}\cline{15-20}
\noalign{\smallskip}
\cline{1-6}\cline{8-13}\cline{15-20}
\noalign{\smallskip}
      \multicolumn{7}{l}
        {\qquad $M_2=1.3 \qquad \alpha =2.0 \qquad$ Pop.$\,${I}} &
      \multicolumn{7}{l}
        {\qquad $M_2=1.6 \qquad \alpha =2.0 \qquad$ Pop.$\,${I}} &
      \multicolumn{6}{l}
        {\qquad $M_2=2.0 \qquad \alpha =2.0 \qquad$ Pop.$\,${I}}\\
\cline{1-6}\cline{8-13}\cline{15-20}
\noalign{\smallskip}
 $P_{\rm orb}^{\rm ZAMS}$ & $P_{\rm orb}^{\rm RLO}$ & $P_{\rm orb}^{\rm f}$ &
 $M_{\rm WD}^{\rm theo}$ & $M_{\rm NS}^{\prime}$ & $t_{\rm X}$ &&
 $P_{\rm orb}^{\rm ZAMS}$ & $P_{\rm orb}^{\rm RLO}$ & $P_{\rm orb}^{\rm f}$ &
 $M_{\rm WD}^{\rm theo}$ & $M_{\rm NS}^{\prime}$ & $t_{\rm X}$ &&
 $P_{\rm orb}^{\rm ZAMS}$ & $P_{\rm orb}^{\rm RLO}$ & $P_{\rm orb}^{\rm f}$ &
 $M_{\rm WD}^{\rm theo}$ & $M_{\rm NS}^{\prime}$ & $t_{\rm X}$\\
\noalign{\smallskip}
\cline{1-6}\cline{8-13}\cline{15-20}
\noalign{\smallskip}
  2.3&0.79& 0.016&0.144&2.44& -- &&  1.5&1.09&  0.08&0.147&2.75& -- & &1.2&1.19&0.05&0.117&2.84& -- \\
 2.35&0.84&  2.30&0.208&2.39&2590&&  1.8&1.15&  1.20&0.189&2.71&2350& &1.3&1.29&2.82&0.211&2.75&1820\\
  2.4&0.96&  5.53&0.229&2.37&1720&&  2.0&1.20&  6.05&0.219&2.68& 780& &1.4&1.39&5.82&0.232&2.75&1070\\
  2.5&1.08&  9.66&0.244&2.36&1140&&  2.1&1.27&  11.2&0.246&2.65& 450& &1.5&1.49&9.56&0.247&2.75& 650\\
  2.6&1.13&  12.8&0.252&2.35& 900&&  2.2&1.30&  19.6&0.263&2.63& 285& &1.6&1.59&14.9&0.259&2.76& 410\\
  2.7&1.17&  15.3&0.257&2.32& 790&&  2.6&1.68&  33.6&0.283&2.13& 136& &1.7&1.69&22.7&0.269&2.76& 218\\
  2.8&1.23&  16.9&0.261&2.28& 725&&  3.0&2.00&  41.8&0.291&2.05& 123& &1.75&1.74&8.1&0.278&2.74& 160\\
  2.9&1.28&  19.0&0.264&2.26& 665&&  4.0&2.90&  58.0&0.303&2.06& 107& &1.77&1.76&0.5&0.281&2.72& 124\\
  3.0&1.35&  21.0&0.267&2.24& 610&&  6.0& 5.0&  87.1&0.320&2.04&78.0& &1.79&1.78&4.1&0.289&2.55& 100\\
  3.4&1.61&  27.9&0.276&2.22& 444&&  8.0& 6.8& 109.6&0.330&2.00&61.0& &1.8&1.79&38.6&0.297&1.83&49.0\\
  4.0&2.02&  36.6&0.285&2.25& 366&& 10.0& 8.7& 133.4&0.339&1.85&42.0& &1.9&1.89&39.7&0.301&1.82&37.8\\
  5.0&3.63&  58.5&0.303&2.20& 190&& 15.0&13.3& 194.7&0.358&1.62&23.2& &2.0&2.00&41.0&0.305&1.81&37.0\\
  6.0&4.89&  73.3&0.311&2.14& 133&& 25.0&22.3& 309.2&0.384&1.50&14.5& &2.2&2.20&44.6&0.310&1.79&37.0\\
 10.0&8.85& 113.5&0.329&2.14&74.0&& 40.0&36.0& 451.5&0.410&1.44& 9.3& &2.4&2.40&48.2&0.313&1.76&36.0\\
 15.0&13.6& 162.9&0.346&1.85&40.0&& 60.0&54.5& 608.0&0.435&1.40& 7.5& &2.6&2.60&51.6&0.316&1.74&34.2\\
 25.0&22.9& 253.1&0.370&1.65&22.6&&  100&92.3& 828.9&0.466$^{*}$&1.37& 5.1& &2.8&2.78&54.7&0.318&1.71&32.8\\
 40.0&37.2& 369.6&0.393&1.54&16.4&&  150& 141&1035.1&0.494$^{*}$&1.36& 4.1& &3.0&2.89&56.2&0.320&1.64&28.5\\
 60.0&56.4& 500.5&0.416&1.47&11.5&&  200& 190&1192.8&0.531$^{*}$&1.35& 3.2& &3.2&2.76&56.4&0.320&1.57&24.0\\
\cline{8-13}
  100&95.4& 714.6&0.448&1.41& 7.6&&     &    &      &     &    &    & &3.4&2.94&59.7&0.322&1.57&24.0\\
  150& 146& 933.5&0.478$^{*}$&1.38& 5.4&&     &    &      &     &    &    & &3.6&3.15&63.3&0.324&1.57&23.8\\
  200& 197&1108.8&0.502$^{*}$&1.36& 4.4&&     &    &      &     &    &    & &3.8&3.37&66.7&0.326&1.57&24.1\\
\cline{1-7}
     &    &      &     &    &    &&     &    &      &     &    &    & &4.0&3.58&69.8&0.327&1.57&24.2\\
     &    &      &     &    &    &&     &    &      &     &    &    & &4.2&3.79&73.0&0.328&1.57&24.0\\
     &    &      &     &    &    &&     &    &      &     &    &    & &$>$4.2& & -- & --  & -- & -- \\
\noalign{\smallskip}
\cline{15-20}
\end{tabular}
\end{center}
\vspace{-1.5cm}
  \begin{list}{}{}
    \item[*]  This is the mass of $M_2$ when the $\sim 0.47\,M_{\odot}$ {He}-core 
              ignites (flash).
    \item[]   $P_{\rm orb}^{\rm ZAMS}$ is the initial orbital period of the
              NS and the unevolved companion.
    \item[]   $P_{\rm orb}^{\rm RLO}$ is the orbital period at onset of RLO. 
    \item[]   $P_{\rm orb}^{\rm f}$ is the final orbital period of the LMXB 
              -- or the (initial) period of the BMSP.
    \item[]   $M_{\rm WD}^{\rm theo}$ is our calculated mass of the helium WD.
    \item[]   $M_{\rm NS}^{\prime}$ is the final mass of the NS if
              $\Delta M_{\rm dp}=0$.
    \item[]   $t_{\rm X}$ is the integrated duration time (Myr) that the binary
              is an {\em active} {X}-ray source.
  \end{list}{}{}
\end{table*}
\setlength{\tabcolsep}{6pt}
\begin{table*}
  \caption{Observed binary pulsars (NS+WD)
           in the Galactic disk. Masses are in units of $M_{\odot}$.}
  \begin{center}
     \begin{tabular}{lclcllllcl}
        \noalign{\bigskip}
        \hline\noalign{\smallskip}
                ${\rm PSR}$ & $P_{\rm orb}$ (days) & $f$ ($M_{\odot}$) 
                  & $M_{\rm WD}^{\rm obs}$
                  & $M_{\rm WD}^{\rm theo}$
                  & $i^{\rm theo}$ (deg.)
                  & $M_{\rm NS}^{\rm max}$
                  & $M_{\rm NS}^{\prime}$
                  & $P_{\rm spin}$ (ms)
                  & Class\\
          \noalign{\smallskip}
          \hline\noalign{\smallskip}
 B0820+02    & 1232 & 0.003 &  0.231 & 0.503 & 26.1 & 6.02 & 1.34 & 
865&{A}$^{*}$\\
 J1803--2712 &  407 & 0.0013    & 0.170 & 0.423 & 22.7 & 7.20 & 1.48 & 334  & 
{A}\\
 J1640+2224  &  175 & 0.0058    & 0.295 & 0.373 & 44.9 & 2.61 & 1.69 & 3.16 & 
{A}\\
 J1643--1224 &  147 & 0.00078   & 0.142 & 0.363 & 21.7 & 7.48 & 1.82 & 4.62 & 
{A}\\
 B1953+29    &  117 & 0.0024    & 0.213 & 0.352 & 33.6 & 3.90 & 1.86 & 6.13 & 
{A}\\
 J2229+2643  & 93.0 & 0.00084   & 0.146 & 0.340 & 23.7 & 6.51 & 1.90 & 2.98 & 
{A}\\
 J2019+2425  & 76.5 & 0.0107    & 0.373 & 0.331 & 73.5 & 1.51 & 1.93 & 3.93 & 
{A}\\
 J1455--3330 & 76.2 & 0.0063    & 0.304 & 0.331 & 53.5 & 2.07 & 1.93 & 7.99 & 
{A}\\
 J1713+0447  & 67.8 & 0.0079    & 0.332 & 0.326 & 61.6 & 1.77 & 1.95 & 4.57 & 
{A}\\
 J2033+1734  & 56.2 & 0.0027    & 0.222 & 0.318 & 38.9 & 3.13 & 1.97 & 5.94 & 
{A}\\
 B1855+09    & 12.33 & 0.00557  & 0.291 & 0.262 & 71.5 & 1.54 & 2.05 & 5.36 & 
{AB}\\
 J1804--2717 & 11.13 & 0.00335  & 0.241 & 0.259 & 54.0 & 2.02 & 2.05 & 9.34 & 
{AB}\\
 J0621+1002  & 8.319 & 0.0271   & 0.540 & 0.251 & $>$90  & 0.51 & 2.06 & 28.9 & 
{C}\\
 J1022+1001  & 7.805 & 0.0833   & 0.872 & 0.249 & $>$90  & 0.18 & 2.06 & 16.5 & 
{C}\\
 J2145--0750 & 6.839 & 0.0242   & 0.515 & 0.245 & $>$90  & 0.54 & 2.07 & 16.1 & 
{C}\\
 J2129--5721 & 6.625 & 0.00105  & 0.158 & 0.244 & 35.4 & 3.48 & 2.07 & 3.73 & 
{AB}\\
 J1603--7202 & 6.309 & 0.00881  & 0.346 & 0.243 & $>$90  & 1.03 & 2.07 & 14.8 & 
{C}\\
 J0437--4715 & 5.741 & 0.00125  & 0.168 & 0.240 & 38.5 & 3.09 & 2.08 & 5.76 & 
{AB}\\
 J1045--4509 & 4.084 & 0.00177  & 0.191 & 0.232 & 46.4 & 2.42 & 2.09 & 7.45 & 
{AB}\\
 J1911--1114 & 2.717 & 0.000799 & 0.143 & 0.222 & 35.2 & 3.48 & 2.09 & 3.63 & 
{B}\\
 J2317+1439  & 2.459 & 0.00221  & 0.206 & 0.220 & 54.8 & 1.97 & 2.10 & 3.44 & 
{B}\\
 J0218+4232  & 2.029 & 0.00204  & 0.201 & 0.216 & 54.1 & 2.00 & 2.10 & 2.32 & 
{B}\\
 B1831--00   & 1.811 & 0.000124 & 0.075 & 0.213 & 18.8 & 8.64 & 2.10 & 521  & 
{B}\\
 J0034--0534 & 1.589 & 0.00127  & 0.169 & 0.211 & 45.0 & 2.50 & 2.11 & 1.88 & 
{B}\\
 J0613--0200 & 1.119 & 0.000972 & 0.154 & 0.205 & 41.5 & 2.78 & 2.12 & 3.06 & 
{B}\\
 B0655+64    & 1.029 & 0.0714   & 0.814 & 0.202 & $>$90  & 0.14 & 2.12 & 196  & 
{C}\\
 J1012+5307  & 0.605 & 0.000580 & 0.128 & 0.193 & 36.1 & 3.33 & 2.13 & 5.26 & 
{B}\\
 B1957+20    & 0.382 & 0.0000052& 0.026 & 0.186 &  7.3 & 35.0 & 2.15 & 1.61 & 
{B}\\
 J0751+1807  & 0.263 & 0.000974 & 0.154 & 0.181 & 48.2 & 2.28 & 2.16 & 3.48 & 
{B}\\
 J2051--0827 & 0.099 & 0.000010 & 0.032 & 0.168 & 10.0 & 21.7 & 2.18 & 4.51 & 
{B}\\
     \hline
     \end{tabular}
  \end{center}
  \bigskip
  \begin{list}{}{}
    \item[]
           The last column gives the classification of the BMSPs.
           Class$\,${A} represents the wide-orbit binaries with {He}-WD
           companions. Class$\,${B} contains the close-orbit binaries
           with {He}-WD companions. In these class$\,${B} systems non-conservative
           angular momentum losses ($\dot{J}_{\rm gwr}$, $\dot{J}_{\rm mb}$
           and irradiation) were dominant in the evolution of the
           progenitor LMXB and $P_{\rm orb}^{\rm i}<P_{\rm bif}$. 
           The subclass {AB} refers to systems in which tidal spin-orbit
           interactions were important but not sufficiently
           strong to finally prevent the orbit from widening
           ($P_{\rm orb}^{\rm i} \approx 2-3$ days). Class$\,${C}
           hosts the BMSPs with heavy {CO}-WD companions. These systems
           evolved through (and survived) a phase with extreme  
           mass-transfer rates and loss of orbital angular momentum
          (e.g. a common envelope).
  \end{list}{}{}
  \bigskip
  \begin{list}{}{}
    \item[]        $M_{\rm WD}^{\rm obs}$ is the mass of the white dwarf
                   assuming $i=60\degr$ (the mean value of a random isotropic
                   distribution) and $M_{\rm NS}=1.4\,M_{\odot}$.
    \item[]        $M_{\rm WD}^{\rm theo}$ is the mass of the white dwarf
                   obtained from our ($P_{\rm orb},M_{\rm WD}$) correlation
                   assuming a Pop.$\,${I+II} composition for the donor.
    \item[]        $i^{\rm theo}$ is the orbital inclination angle required for 
                   $M_{\rm WD}^{\rm obs}=M_{\rm WD}^{\rm theo}$;
                   $i>90\degr$ means there is no solution 
                   (the observed WD is `too heavy').
    \item[]        $M_{\rm NS}^{\rm max}$ is the maximum value for $M_{\rm NS}$
                   obtained from the observed binary mass function, $f$ 
                   using $i=90\degr$ and $M_{\rm WD}=M_{\rm WD}^{\rm theo}$.
    \item[]        $M_{\rm NS}^{\prime}$ is a rough estimate of the 
                   potential maximum post-accretion mass of the NS
                   assuming $\Delta M_{\rm dp}=0$ (see curves in Fig.~4b).

    \item[]        For references to the observed values ($P_{\rm orb}$, $f$ and
                   $P_{\rm spin}$) see e.g. Camilo (1995) and Lorimer~et~al (1996).
    \item[*]       The companion of this very wide-orbit pulsar might be a 
                   {CO} white dwarf (cf. Section~6.2).
  \end{list}
  \bigskip
  \bigskip
  \bigskip
\end{table*}

\end{document}